\documentclass[aps,prl,reprint,amssymb,showpacs,superscriptaddress,notitlepage,onecolumn]{revtex4-2}
\usepackage{bm}
\usepackage{graphicx,hyperref}
\usepackage{dcolumn}
\usepackage{braket}
\usepackage{natbib}
\usepackage{amsmath}
\usepackage{color}
\usepackage{siunitx}
\usepackage{mathtools,amssymb}
\usepackage{soul}
\usepackage{float}
\usepackage{amssymb}
\usepackage{esint}
\usepackage{mathtools}
\usepackage{physics}
\usepackage{makecell}
\usepackage{multirow}
\usepackage{array}
\usepackage{geometry}
\usepackage{url}
\usepackage{adjustbox}
\usepackage{rotating}
\usepackage{listings}
\usepackage{threeparttable,booktabs}
\usepackage{fontawesome}
\usepackage{cleveref}
\usepackage{svg}
\usepackage{ragged2e}

\begin{document}
\title{Excitonic Mott insulator in a Bose-Fermi-Hubbard system  of moir\'e $\rm{WS}_2$/$\rm{WSe}_2$ heterobilayer}

\author{Beini Gao}
\altaffiliation{These authors contributed equally to this work}
\affiliation{Joint Quantum Institute (JQI), University of Maryland, College Park, MD 20742, USA}
\author{Daniel G. Suárez-Forero$^*$}
\email{dsuarezf@umd.edu}
\affiliation{Joint Quantum Institute (JQI), University of Maryland, College Park, MD 20742, USA}
\author{Supratik Sarkar}
\altaffiliation{These authors contributed equally to this work}
\affiliation{Joint Quantum Institute (JQI), University of Maryland, College Park, MD 20742, USA}
\author{Tsung-Sheng Huang}
\affiliation{Joint Quantum Institute (JQI), University of Maryland, College Park, MD 20742, USA}
\author{Deric Session}
\affiliation{Joint Quantum Institute (JQI), University of Maryland, College Park, MD 20742, USA}
\author{Mahmoud Jalali Mehrabad}
\affiliation{Joint Quantum Institute (JQI), University of Maryland, College Park, MD 20742, USA}
\author{Ruihao Ni}
\affiliation{Department of Materials Science and Engineering, University of Maryland, College Park, MD 20742, USA}
\author{Ming Xie}
\affiliation{Condensed Matter Theory Center, University of Maryland, College Park, MD 20742, USA}
\author{Pranshoo Upadhyay}
\affiliation{Joint Quantum Institute (JQI), University of Maryland, College Park, MD 20742, USA}
\author{Jonathan Vannucci}
\affiliation{Joint Quantum Institute (JQI), University of Maryland, College Park, MD 20742, USA}
\author{Sunil Mittal}
\affiliation{Joint Quantum Institute (JQI), University of Maryland, College Park, MD 20742, USA}
\author{Kenji Watanabe}
\affiliation{National Institute for Materials Science, Tsukuba, Japan}
\author{Takashi Taniguchi}
\affiliation{National Institute for Materials Science, Tsukuba, Japan}
\author{Atac Imamoglu}
\affiliation{Institute of Quantum Electronics, ETH Zurich, CH-8093 Zurich, Switzerland}
\author{You Zhou}
\affiliation{Department of Materials Science and Engineering, University of Maryland, College Park, MD 20742, USA}
\affiliation{Maryland Quantum Materials Center, College Park, Maryland 20742, USA}
\author{Mohammad Hafezi}
\email{hafezi@umd.edu}
\affiliation{Joint Quantum Institute (JQI), University of Maryland, College Park, MD 20742, USA}
\affiliation{Institute for Theoretical Physics, ETH Zurich, 8093 Zurich, Switzerland}

%%%%%%%%%%%%%%%%%%%%%%%%%%%%%%%%%%%%%%%%%%%%%%%%%%%%%%%%%%%%%%%%%%%%%%%%%%%%%%%%%%%%%%%%%%%%%%%%%%%%%%%%%%%%%%%%%%%%%%%%%%%%%%

\begin{abstract}

Understanding the Hubbard model is crucial 
for investigating various quantum many-body states and its fermionic and bosonic versions have been largely realized separately. Recently, transition metal dichalcogenides heterobilayers have emerged as a promising platform for simulating the rich physics of the Hubbard model. In this work, we explore the interplay between fermionic and bosonic populations, using a $\rm{WS}_2$/$\rm{WSe}_2$ heterobilayer device that hosts this hybrid particle density. We independently tune the fermionic and bosonic populations by electronic doping and optical injection of electron-hole pairs, respectively. This enables us to form strongly interacting excitons that are manifested in a large energy gap in the photoluminescence spectrum. The incompressibility of excitons is further corroborated by observing a suppression of exciton diffusion with increasing pump intensity, as opposed to the expected behavior of a weakly interacting gas of bosons, suggesting the formation of a bosonic Mott insulator. We explain our observations using a two-band model including phase space filling. Our system provides a controllable approach to the exploration of quantum many-body effects in the generalized Bose-Fermi-Hubbard model.
\end{abstract}

%%%%%%%%%%%%%%%%%%%%%%%%%%%%%%%%%%%%%%%%%%%%%%%%%%%%%%%%%%%%%%%%%%%%%%%%%%%%%%%%%%%%%%%%%%%%%%%%%%%%%%%%%%%%%%%%%%%%%%%%%%%%%%

\maketitle
\subsubsection{\bf Introduction}
The rich physics of the Hubbard model has brought fundamental insights to the study of many-body quantum physics \cite{hubbard1693}. Initially proposed for electrons on a lattice, different fermionic and bosonic versions of this model have been simulated in various platforms, ranging from ultracold atoms \cite{Greiner2002} to superconducting circuits \cite{Carusotto2020}. Recently, bilayer transition metal dichalcogenides (TMDs) have become a versatile platform to study the Hubbard model thanks to the coexistence of several intriguing features such as the reduction of electron hopping due to the formation of moir\'e lattice with large lattice constant, and the existence of both intra- and inter-layer excitons. These characteristics have enabled the realization of numerous effects of many-body physics such as metal-to-Mott insulator transition \cite{Ghiotto2021,Li2021,Shimazaki2020,Tang2020,Zhang2020,Wang2020}, generalized Wigner crystals \cite{Xu2020,Huang2021,Regan2020,Li2021-4,Li2021-5},
exciton-polaritons with moir\'{e}-induced nonlinearity \cite{Zhang2021}, stripe phases \cite{Jin2021}, light-induced ferromagnetism \cite{Wang2022}. Moreover, there have been recent exciting perspectives of exploring such effects in light-matter correlated systems\cite{Carusotto2013,Carusotto2020,Bloch2022}. While typically the fermionic and bosonic versions of the Hubbard model are explored independently, combining these two models in a single system holds intriguing possibilities for studying mixed bosonic-fermionic correlated states \cite{Xiong2023,Park2023}. 

In this work, we demonstrate Bose-Fermi-Hubbard physics in a TMD heterobilayer. We independently control the population of fermionic (electronic) particles by doping with a gate voltage ($V_{\rm g}$), and the population of bosonic (excitonic) states by pumping with a pulsed optical drive of intensity $I$. Harnessing these two control methods, we realize strongly interacting excitons. In particular, we show the incompressibility of excitonic states near integer filling by observing an energy gap in photoluminescence, accompanied by an intensity saturation. Remarkably, we observe the suppression of diffusion, a strong indication of the formation of a bosonic Mott insulator of excitons.

\begin{figure}
    \centering    
    \includegraphics[width=\columnwidth]{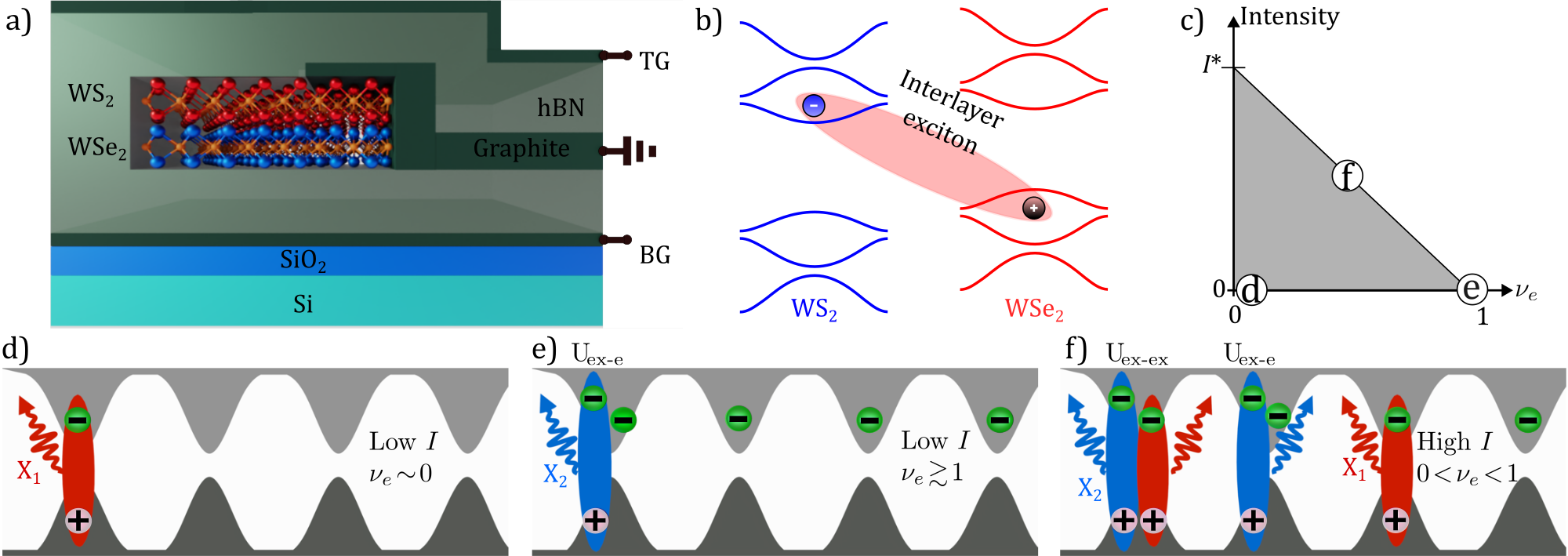}
    \caption{
    (a) Schematic of the $\text{WS}_2$/$\text{WSe}_2$ dual-gate device. The TMD heterobilayer is embedded between two symmetric gates: top gate (TG) and bottom gate (BG). (b) Depiction of the type-II band alignment of the bilayer. The blue and red curves denote bands from $\text{WS}_2$ and $\text{WSe}_2$, respectively. The shaded ellipse indicates the formation of interlayer excitons composed of an electron from the $\text{WS}_2$ conduction band and a hole from the $\text{WSe}_2$ valence band. (c) Phase diagram of the system. The population of the moir\'e lattice can be controlled via two independent parameters: the gate voltage changes the electronic filling factor ($\nu_e$), and the optical pump creates a population of excitons, proportional to the input intensity. In the gray area, the system behaves as a mixed gas of bosonic and fermionic particles. As one approaches the upper limit (black line), the system becomes incompressible  due to the saturation of single occupancy states. (d-f) Interlayer exciton formation under optical excitation for three different regimes governed by the pump intensity ($I$) and $\nu_e$: (c) low $I$ and $\nu_e \sim 0$, (d) low $I$ and $\nu_e \sim 1$, (e) high $I$ and $ 0 < \nu_e < 1$. $\text{X}_1$ ($\text{X}_2$) denotes PL emission from singly (doubly) occupied moir\'e lattice sites. $\text{X}_2$ can originate from either electron-exciton (U$_{\rm{ex-e}}$) or exciton-exciton (U$_{\rm ex-ex}$) double occupancies.
    }
    \label{fig1}
\end{figure}

\subsubsection{\bf Results}
\noindent\textbf{Physical system and experimental design}\\
To demonstrate these effects, we use a moir\'e lattice created by stacking two monolayers of $\rm{WS}_2$ and $\rm{WSe}_2$, with symmetric top and bottom gates. Fig.~\ref{fig1}a shows a schematic illustration of the heterobilayer device (See Supplementary Note 1 for details). Due to the type-II band alignment of the heterostructure (Fig.~\ref{fig1}b), negative doping results in a population of electrons in the WS$_2$ subject to the moir\'e potential of the bilayer. The ratio between the density of this population and the density of moir\'e sites in the structure determines the so-called electronic filling factor ($\nu_e$). The optical pump results in the formation of an energetically favorable interlayer exciton (X) \cite{Xu2018}, by pairing between an electron in WS$_2$ and a hole in WSe$_2$ (represented in Fig.~\ref{fig1}b). In order to explore different regimes of Bose-Fermi-Hubbard model, we control the bosonic and fermionic populations by changing $I$ and $V_{\rm g}$, respectively. This can be compared to ultra-cold atom implementation of Bose-Fermi mixture where the respective populations are fixed in each experiment \cite{gunter2006bose}.  Before discussing our experimental observation, we discuss three limiting cases that determine the phase space of our system, as indicated in Fig.~\ref{fig1}c. The corresponding physical scenarios are represented in panels d to f. First, in the weak excitation limit and low electronic filling factor ($\nu_e\!\sim\!0$) regime, the system's photoluminescence (PL) emission originates exclusively from the few X states in the quasi-empty lattice (panel d). This emission comes from excitons in lattice sites where they are the only occupant particles, namely, ``single occupancy states" ($\text{X}_1$). Upon increasing $\nu_e$, the number of singly occupied sites decreases, and in the limiting case of $\nu_e\!\geq\!1$, as represented in panel e, the optically generated excitons can only form in lattice sites already occupied by charged particles. In this case, the required energy to form the exciton increases due to the on-site Coulomb repulsion, and hence the PL emission has new branch with higher energy than the previous regime. Consequently, the PL originates from lattice sites with an electron-exciton double occupancy ($\text{X}_2$). Finally, we consider the case where the electronic doping is below the threshold required to reach a fermionic Mott insulator ($0\!<\!\nu_e\!<\!1$) but $I$ is strong enough to optically saturate the single occupancy states. The extra excitons create a number of sites with electron-exciton or exciton-exciton double occupancies (panel f). In this case, the PL emission corresponds to mixed contributions from exciton-exciton and exciton-electron interaction (U$_{\text{ex-ex}}$ and U${_\text{ex-e}}$); the individual peaks cannot be distinguished in a single spectrum due to the broadness of linewidths.
%, because the broadness of linewidths does not allow one to distinguish individual peaks in a single spectrum.} 
Therefore, in this regime, the emitted light is only a combination of the $\text{X}_1$ and $\text{X}_2$ PL emission.  This interplay between exciton and electron occupancy can lead to situations in which the moir\'e lattice is completely filled with a mixed population of fermions and bosons, forming a hybrid incompressible state. Specifically, in the limit of weak electronic tunneling, excitons can form a Mott insulating state, in the remainder of sites that are not filled by electronic doping. Note the line in Fig.~\ref{fig1}c denoting panel f is an asymptote since optical pumping can not fully saturate an exciton line. At $\nu_e\!=\!0$, this intensity is denoted as $I^*$.  (See Supplementary Note 8 for details).

%%%%%%%%%%%%%%%%%%%%%%%%%%%%%%%%%%%%%%%%%%%%%%%%%%%%%%%%%%%%%%%%%%%%%%%%%%%%%

\noindent\textbf{System's properties for varying electronic (fermionic) occupation}\\
To experimentally investigate these regimes, we perform PL measurements, with varying pump power and backgate voltage. A detailed description of the optical setup can be found in Supplementary Note 2. We use pulsed excitation to achieve high exciton density while reducing thermal effects by keeping low average power. Experiments with  CW excitation are consistent with the presented data, as shown in the Supplementary Note 6. Fig.~\ref{fig2}a-c shows the PL dependence at three different intensities as schematically shown in panel d. Fig.~\ref{fig2}a shows the normalized doping-dependent PL spectrum for low $I$ (0.08$\rm{\mu W} /\rm{\mu m}^2$), which corresponds to low bosonic occupation. The fermionic occupation $\nu_e$ is varied between $0$ and $1.1$. For low $\nu_e$, PL emission is detected only from $\rm{X}_1$. However, at $\text{V}_g\approx2.98$V, we detect a transition in the PL emission to X$_2$. This transition corresponds to the formation of X's in the presence of an incompressible fermionic Mott insulator \cite{Miao2021,Bai2022}. From the reflectivity measurement and calculations from a capacitor model, we attribute $\text{V}_g\!=\!2.98$ V to $\nu_e\!=\!1$ (see Supplementary Note 3). The energy gap between $\text{X}_1$ and $\text{X}_2$ is $\Delta E\approx29 \text{meV}$, which corresponds to the on-site Coulomb repulsion energy between an electron and an exciton (U$_{\rm ex-e}$). We elaborate on this energy gap later in the sub-section ``Energy map along the phase space''. The dim mid-gap features between X$_1$ and X$_2$ at $\nu_e\!\sim\!1$ are strongly position dependent and disappear at higher power. This indicates that such emission is from localized excitons. Fig.~\ref{fig2}b shows the PL spectrum under pump intensity equal to 12.1 $\mu W/\mu m^2$. It is worth noticing that the $V_{\rm g}$ at which the PL signal from $\rm{X}_2$ is detected, is lower than in panel a. The system is therefore in the regime depicted in Fig.~\ref{fig1}f. Upon further increasing the pump intensity $\rm{X}_2$ can be detected even at $\nu_e=0$, as observed in Fig.~\ref{fig2}c. In this case, the PL emission originates solely from double occupancy of excitons in a moir\'e lattice site, suggesting that, for high $I$, purely bosonic states of strongly interacting excitons are created. Comparing Fig.~\ref{fig2}a and Fig.~\ref{fig2}c, one can observe that in the former case, the emergence of the X$_2$ peak corresponds to a sharp suppression of X$_1$, while in the latter case, both peaks coexist.
%A noteworthy feature in Fig.~\ref{fig2}a is the emergence of the X$_2$ peak combined with a sharp suppression of the X$_1$ peak, unlike in Fig.~\ref{fig2}c, where there is a coexistence of both peaks. 
This indicates the nature of the double occupancy: in the first scenario, the exciton is forming in the presence of an electron, and after its recombination, there are no other optical excitations in the system. In contrast, the coexistence of both peaks in panel c shows that upon double exciton occupancy, the recombination of X$_2$ precedes the recombination of X$_1$.

\begin{figure}
    \centering    
\includegraphics[width=\columnwidth]{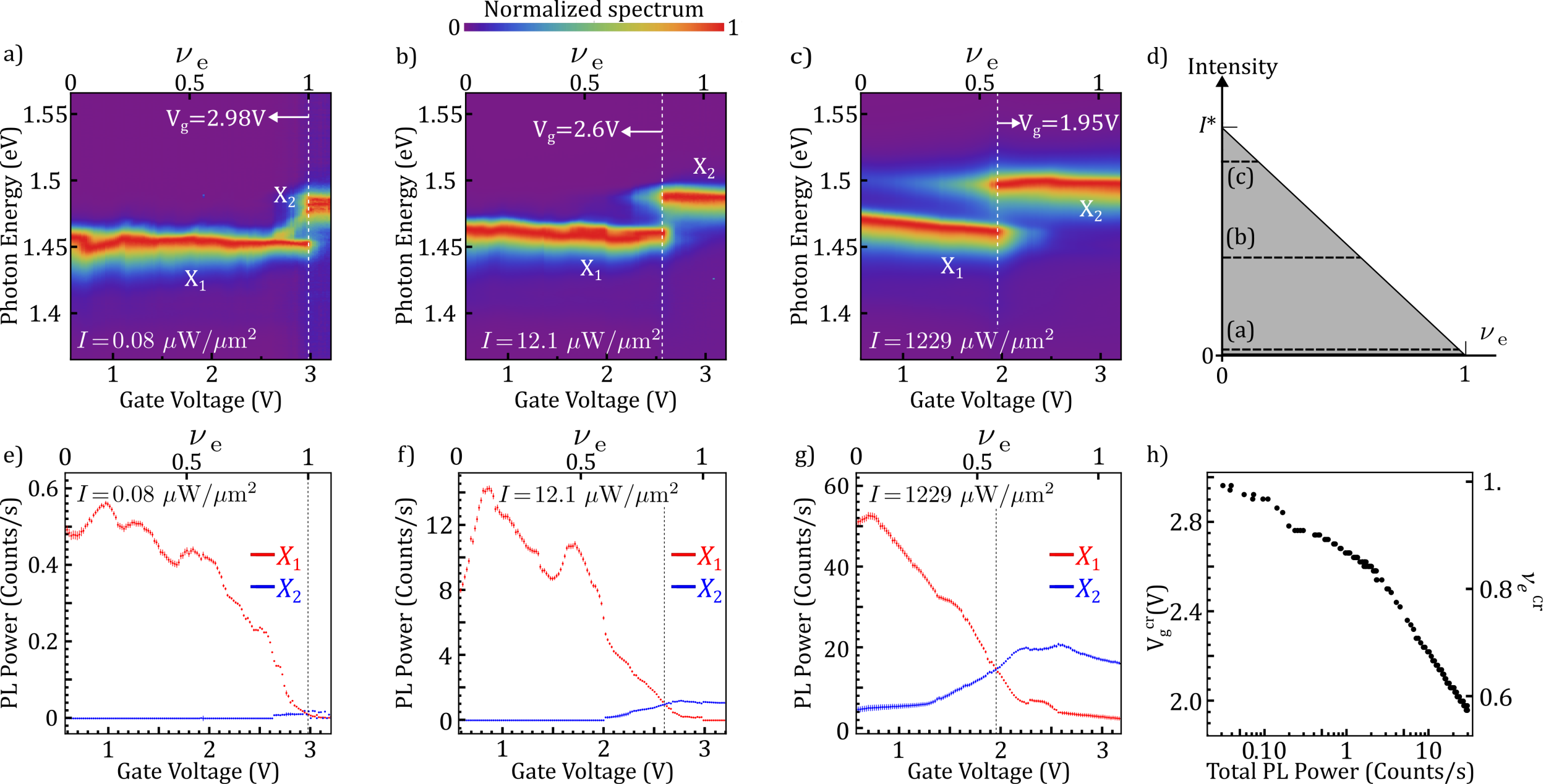}
    \caption{(a-c) Normalized PL spectrum as a function of gate voltage ($\nu_e$) for three different pump intensities: $\text{I}=0.08$ $\rm{\mu W\mu m^2}$ (a), $\text{I}=12.1$$\rm{\mu W/\mu m^2}$ (b) and $\text{I}=1229$$\rm{\mu W/\mu m^2}$ (c). The peaks associated with single (X$_1$) and double (X$_2$) occupancy are indicated on each panel. The dashed lines indicate the gate voltages at which the PL intensity X$_2$ exceeds X$_1$. The dashed black lines of panel (d) indicate the measurement ranges of (a-c). (e-f) Evolution of the PL intensity for X$_1$ (red) and X$_2$ (blue) as a function of gate voltage for the same values of pump intensities displayed in panels (a-c). The electron filling factor at which X$_2$ exceeds X$_1$ decreases as pump intensity increases. Panel (h) shows the gate voltage at which the intensity of X$_2$ exceeds that of X$_1$, as a function of the total PL intensity. The error bars represent the standard errors for the parameter estimates in the fitting routine.
    } 
    \label{fig2}
\end{figure}

From the observation described in the previous paragraph, we conclude that the detection of PL emission with $\rm{X}_1$ and $\rm{X}_2$ energies benchmarks the formation of exciton states in singly and doubly occupied lattice sites, respectively. At $\nu_e\!=\!0$, the X$_1$ peak in Fig.~\ref{fig2}c is blueshifted with respect to Fig.~\ref{fig2}a. We associate this feature with a mean-field effect due to exciton-exciton interaction. As we increase the electronic doping, fewer sites are available to create X$_1$ excitons and on those occupied sites, only X$_2$ is created. Consequently, the effective population of X$_1$ excitons is decreased. Therefore the mean-field shift is suppressed to the point that at high filling ($\nu_e\!\sim\!1$) the X$_1$ energy is the same as in the case of low pump intensity.
Next, in order to understand the interplay between fermionic and bosonic lattice occupancies in each regime, we perform a quantitative analysis of their respective integrated intensity. We extract these values from the collected PL spectra using a computational fitting method (see Supplementary Note 7 for further details). Fig.~\ref{fig2}e-g displays this intensity dependence on $\nu_e$ for the same $I$ range of panels a-c. We notice that as electrons fill the system’s phase space (upon increasing $V_g$), the number of accessible single-occupancy states decreases. As a consequence, the integrated intensity of X$_1$ reduces with increasing $\nu_e$. Remarkably, for each intensity, there is a critical $\nu_e$ after which the PL emission of $\rm{X}_2$ exceeds that of $\rm{X}_1$. The gate voltage at which the crossing takes place ($V_{\rm g}^{\rm cr}$) is highlighted on each panel by a vertical dashed line. This line indicates a constant ratio between the X$_1$ and X$_2$ populations. The crossing takes place at lower $\nu_e$ upon increasing $I$, as expected. In Fig.~\ref{fig2}h, we track $V_{\rm g}^{ \rm cr}$ as a function of the total collected PL emission, which gives an indication of the total number of excitons in both $\rm{X}_1$ and $\rm{X}_2$ branches. We observe a clear trend: a higher total population of excitons results in a faster saturation of the single occupancy states and hence an increasing number of double occupancy states.\\

%%%%%%%%%%%%%%%%%%%%%%%%%%%%%%%%%%%%%%%%%%%%%%%%%%%%%%%%%%%%%%%%%%%%%%%%%%%%%%%%%%%%%%%%%%%%%%%%%%%%%%%%%%%%%%%%%%%%%%%%%%%%%%

\begin{figure}
    \centering    
    \includegraphics[width=\columnwidth]{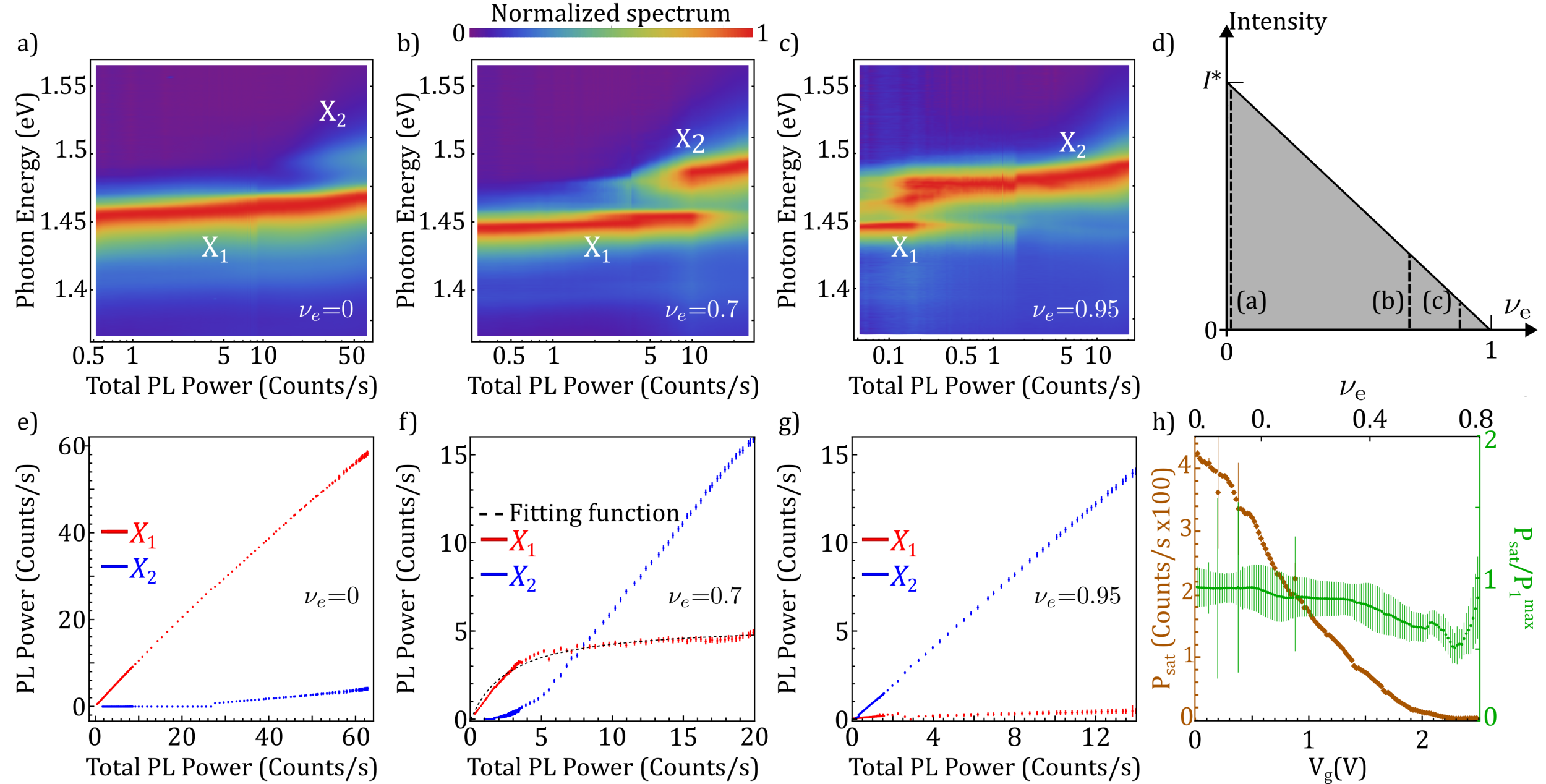}
    \caption{(a-c) Normalized PL spectrum as a function of the total collected PL power for three different electronic filling factors. The peaks associated with single (X$_1$) and double (X$_2$) occupancy are indicated on each panel. Panel (d) indicates the ranges of $I$ and $\nu_e$ for the measurements shown in panels (a-c). (e-f) evolution of the PL power for X$_1$ (red) and X$_2$ (blue) as a function of the total collected PL power for the same values of $\nu_e$ displayed in panels (a) to (c). Panel (f) displays the fitting function (dashed black line) employed to extract $\text{P}_{\text{sat}}$ and $\text{P}_1^\text{max}$ (described in the text). (h) Evolution of $\text{P}_{\text{sat}}$ (brown) as a function of the gate voltage ($\nu_e$). As expected from our phase-space filling model, its value reduces with increasing filling factor. The quantity $\text{P}_{\text{sat}}/\text{P}_1^{\text{max}}$ (green) shows good agreement with the theoretical model. The error bars represent the standard errors for the parameter estimates in the fitting routine.}
    \label{fig3}
\end{figure}

\noindent\textbf{System's properties for varying excitonic (bosonic) occupation}\\
Next, in order to trace the role of the optical pump and the optical saturation that leads to the formation of incompressible bosonic states, we investigate the PL for varying $I$ for different $\nu_e$. In Fig.~\ref{fig3}a-c, we focus on three different values of $\nu_e$, as indicated in panel d, and study the PL spectrum for increasing emitted PL power. For zero fermionic occupancy (panel a), X$_2$ contributes to the emission only at very high total PL emission intensity. In panels b and c, we increase the electronic doping to $\nu_e\!=\!0.7$ and $\nu_e\!=\!0.95$, respectively, and as expected, the total PL at which we detect X$_2$ decreases. In the low power region, panel c shows the PL emission from mid-gap states also observed in Fig.~\ref{fig2}a. Apart from the energy gap in the emission, we observe a blueshift of the X$_1$ line with increasing PL power. Assuming the weak tunneling regime, this shift should be equal to $U_{\rm ex-ex} \langle \hat{x}^\dagger \hat{x} \rangle$, where $\hat{x}^\dagger$ is the creation operator of an exciton. For example, in Fig.~\ref{fig3}b for total PL power at 2 counts/s, the bosonic occupation is $\langle \hat{x}^\dagger \hat{x} \rangle \simeq 0.2$. This corroborates with the energy gap that occurs at 10 counts/s for an estimated unity filling ($\langle \hat{x}^\dagger \hat{x} \rangle \simeq 1$). We present a fully quantum theoretical analysis of this observation in Supplementary Note 10. Panels e-g show the intensities of X$_1$ and X$_2$ for the values of $\nu_e$ in panels a-c. As expected, in panel e, one can observe that the intensity of the X$_1$ PL emission increases monotonically, and it starts to saturate only at very high total PL emission regimes. Upon filling the moir\'e lattice with one exciton or one electron per site, the X$_1$ PL intensity saturates. With higher $\nu_e$, the saturation occurs at lower $I$, as shown in panels f and g. Since this saturation corresponds to filling the single occupancy states, we associate it with the establishment of an incompressible bosonic Mott insulator. Note that this bosonic Mott insulator is in a drive-dissipative regime, similar to the demonstration in superconducting qubit systems \cite{ma2019dissipatively}.

To quantitatively analyze this saturation effect, we fit the X$_1$ PL power (P$_1$) to the function P$_1=\!\text{P}_1^{\text{max}}\frac{{\rm P}}{\text{P}+\rm{P}_{\rm sat}}$, where P is the total PL power. From the fitting, we extract P$_1^{\rm max}$ which is the asymptotic value of the X$_1$ emitted PL power, and P$_{\rm sat}$ which determines the total PL of saturation. This functional form corresponds to the expected system behavior when the charge gap U is sufficiently large to permit the utilization of a phase-space filling model to treat both single and double occupancy states (details in Supplementary Note 8). Fig.~\ref{fig3}f includes an example of the fitting function (dashed black line). According to our model, the value of $\text{P}_{\text{sat}}$ should decrease with increasing $\nu_e$ because a lower excitonic population is required to achieve the incompressible states. The compiled data for the full range of $\nu_e$, shown in panel h with brown marks, is in good agreement with the expected trend. From the model, we can also infer that the quantity $\text{P}_{\text{sat}}/\text{P}_1^{\text{max}}$ should be independent of the electronic doping level because both quantities depend linearly on $1-\nu_e$; higher electronic occupancy implies less single occupancy states available to host an exciton. The green marks in Fig.~\ref{fig3}h represent this behavior, which is in good agreement with the model. We conclude that the saturation of single occupancy states is directly reflected in the intensity of X$_1$, enabling the extraction of the conditions under which the incompressible states occur. Importantly, this enables a direct calibration of the bosonic and fermionic fractions in the system.\\

%%%%%%%%%%%%%%%%%%%%%%%%%%%%%%%%%%%%%%%%%%%%%%%%%%%%%%%%%%%%%%%%%%%%%%%%%%%%%%%%%%%%%%%%%%%%%%%%%%%%%%%%%%%%%%%%%%%%%%%%%%%%%%

\begin{figure}
    \centering    
    \includegraphics[width=0.95\columnwidth]{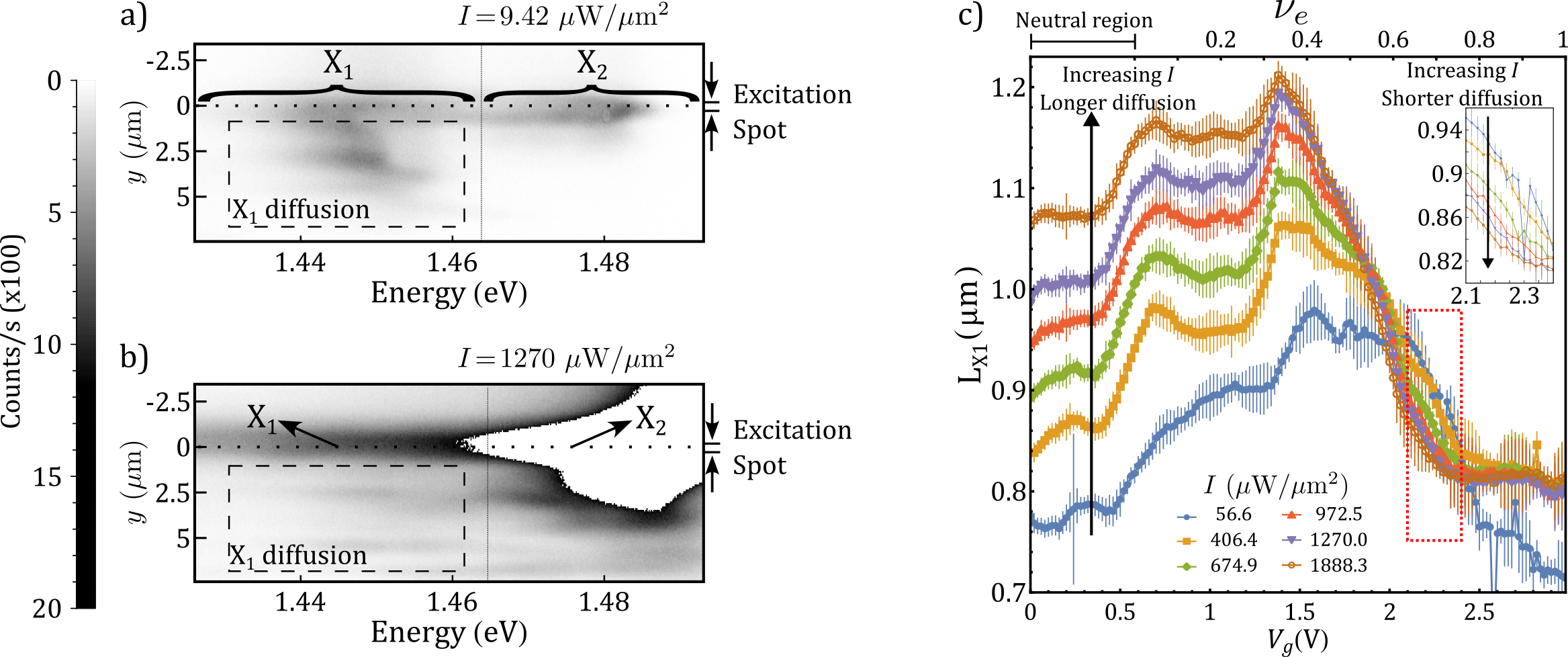}
    \caption{Spectrally and spatially-resolved diffusion pattern at $\nu_e\!=\!0.73$ $({\rm V_g}=2.34$V) for low (a) and high (b) $I$. The dashed rectangle highlights the region where the suppression of diffusion can be observed.
    (c) Exciton diffusion length as a function of the gate voltage for a range of $\nu_e$ and for different input intensities. For low $\nu_e$ the diffusion length increases with $I$ due to exciton repulsive interaction. Upon further filling the moir\'e lattice, the trend inverts, indicating the optical realization of incompressible states. The inset is a zoom-in of the red dotted rectangle to highlight the reduction of $L_{X1}$ with increasing $I$. The error bars represent the standard errors for the diffusion length estimated from the exponential fitting.}
    \label{fig4}
\end{figure}

\noindent\textbf{Exciton diffusion measurements}\\
In order to further validate the incompressible nature of excitonic states, we perform diffusion measurements of the interlayer excitons \cite{Jiang2021}. For a steady population of excitons created by a continuous-wave laser pump, the diffusion length carries information about the nature of the state: an incompressible bosonic state is expected to have a lower diffusion length than a weakly interacting one. We spatially image the diffusion pattern with spectral resolution and extract the diffusion length ($\rm{L}_{X1}$) of the single occupancy excitons. The choice of L$_{X1}$ as an appropriate quantity to benchmark the incompressibility of bosonic Mott insulating states, assumes a constant exciton lifetime with varying population. This is supported by previous reports in the literature that show the independence of this quantity over three orders of magnitude of pumping power \cite{Jauregui2019}. The downward diffusion image has patterns that originate in the inhomogeneous surface of the bilayer. Although the inhomogeneities on that side hinder the extraction of $L_{X1}$, the optically-induced suppression of the diffusion length for constant $\nu_e$ can be clearly observed in this region (Fig.~\ref{fig4}a-b). The population injected at $y=0$ (dotted line) propagates, and its emission pattern is monitored along a range of $5$ $\mu m$ (dashed rectangle). The color scale is the same for both panels. Panel b shows a reduction of the diffused X$_1$ population in comparison to panel a. For the quantitative analysis of this observation, it is necessary to use a fitting routine, for which the smooth pattern on top of the injection point ($y\!<\!0$) is more reliable. Figure \ref{fig4}c shows the extracted L$_{{\rm X}1}$ as a function of $V_{\rm g}$ for different pump intensities from the exponentially decaying spatial diffusion pattern in this region. We provide more details about the analysis of the diffusion data in Supplementary Note 9. For low electronic density, the exciton diffusion length increases as the power is augmented. This trend, highlighted by the upward arrow, is in agreement with the expected behavior for weakly interacting bosons \cite{Jauregui2019,Unuchek2019}. Remarkably, as the electronic filling factor increases, the trend completely inverts (inset). This is a direct signature of the bosonic Mott insulator formation since the bulk is incompressible and the melting only occurs at the edge.\\

\begin{figure}
    \centering    
    \includegraphics[width=\columnwidth]{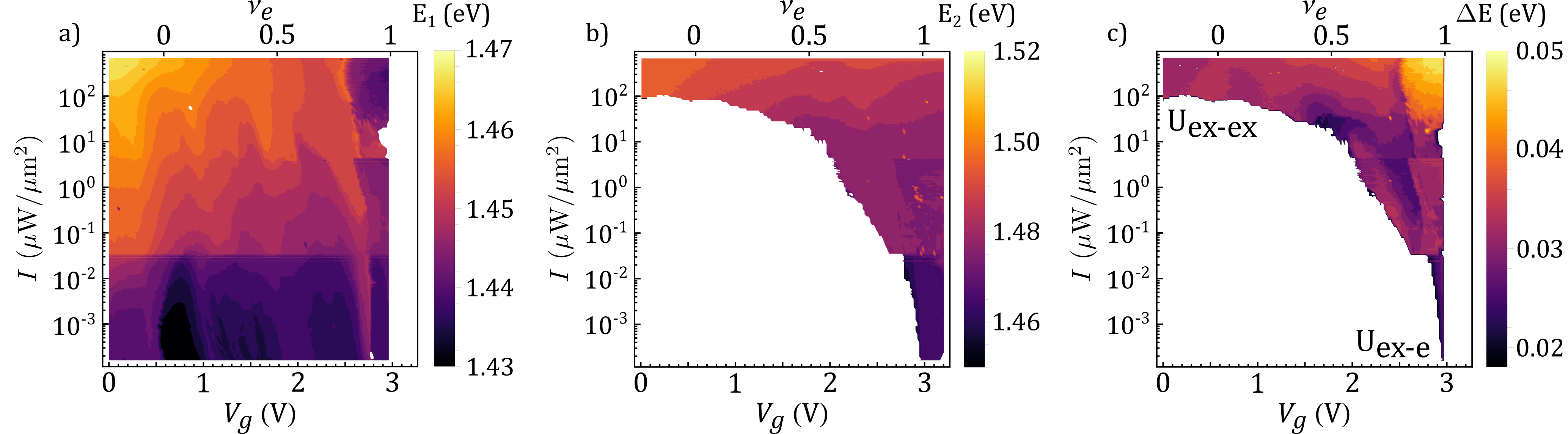}
    \caption{Energy of the X$_1$ (a) and X$_2$ (b) PL emission as a function of gate voltage and pump intensities. The white areas correspond to the range of parameters where the corresponding peak completely vanishes. When X$_1$ and X$_2$ coexist, we extract the energy difference, as shown in panel (c). }
    \label{fig5}
\end{figure}

\noindent\textbf{Energy map along the phase space}\\
The implemented fitting algorithm allows us to track the changes in the energy of both species of excitons and the energy gap between them. These results are presented in Fig.~\ref{fig5}. Panels a and b show the central energies of the peaks X$_1$ and X$_2$ in the space of parameters for which each peak is detectable. In the range where both of them can be detected, their energy difference $\Delta E$ (panel c) provides important information about the nature of the interactions taking place in the system. 
In the case of low electronic occupancy and high exciton density (top left corner of the panel), $\Delta E$ corresponds to the exciton-exciton interaction gap (U${\text{ex-ex}}\!\sim\!32$ meV). Conversely, at high $\nu_e$ and low exciton density (bottom right corner), this gap depends on the exciton-electron interaction (U${\text{ex-e}}\!\sim\!27$ meV). The gradual change in the nature of the interactions taking place in the system along the parameters space is reflected in the change of $\Delta E$. Interestingly, the largest energy gap takes place for states with high occupation of bosons and fermions (top right corner), which is consistent with a blueshift of the X$_2$ PL peak due to the high population of excitons with large Bohr radius repelling through dipolar interaction.
%%%%%%%%%%%%%%%%%%%%%%%%%%%%%%%%%%%%%%%%%%%%%%%%%%%%%%%%%%%%%%%%%%%%%%%%%%%%%%%%%%%%%%%%%%%%%%%%%%%%%%%%%%%%%%%%%%%%%%%%%%%%%%

\subsubsection{\bf Discussion}
In summary, we demonstrated a Mott insulating state of excitons in a hybrid Bose-Fermi Hubbard system formed in a TMD heterobilayer. While our incompressibility observation was based on spatially resolved diffusion in the steady-state limit, one can explore interesting non-equilibrium physics due to the relatively long lifetime of interlayer excitons. More generally, spatiotemporally resolved measurements, combined with independent tunability of fermionic and bosonic populations, make it possible to investigate the equilibrium and non-equilibrium physics of Bose-Fermi mixtures. Moreover, a quantum microscopic model capable of fully describing such a driven-dissipative Bose-
Fermi mixture remains an open area of research. The novel experimental diffusion method used to benchmark the excitonic incompressibility opens exciting perspectives for the simulation of complex dynamics in many-body quantum systems that range from a single bosonic particle in a Fermi sea to a strongly interacting gas of bosons. Particularly intriguing examples are the optical investigation of charge and spin physics in integer and fractional fillings, e.g., Mott excitons \cite{huang,Zhang2022-3} or spin liquids \cite{Moore_CSL_2020,Knap_CSL_2022,Rademaker_CSL_TMD,Dominik2022}, and fractional Chern insulators \cite{FCI-moire2021,crepel2022anomalous}.

%During the course of this project, the authors became aware of other works exploring the interaction between interlayer excitons in similar systems \cite{Xiong2023,Park2023}.

\subsubsection{\bf Methods}
\textbf{Device fabrication}:
The $\rm{WSe}_2/\rm{WS}_2$ heterostructure was fabricated using a dry-transfer method with a stamp made of a poly(bisphenol A carbonate) (PC) layer on polydimethylsiloxane (PDMS). All flakes were exfoliated from bulk crystals onto Si/SiO$_2$ (285 nm) and identified by their optical contrast. The top/bottom gates and TMD contact are made of few-layer graphene. The PC stamp and samples were heated to $60^{\circ}$C during the pick-up steps and released from the stamp to the substrate at $180^{\circ}$C. The PC residue on the device was removed in chloroform followed by a rinse in isopropyl alcohol and ozone clean. Sample transfer was performed in an argon-filled glovebox for improved interface quality. The electrodes consist of 3.5 nm of chromium and 70 nm of gold. They were fabricated using standard electron-beam lithography techniques and thermal evaporation.

\textbf{Optical measurements:}
The sample is kept in a dilution refrigerator at a temperature of $3.5$K. For PL measurements, we use a confocal microscopy setup. Our pumping source is a pulsed Ti:Sapphire laser tuned at $720$nm ($1.722$eV), with a pulse duration of $100$fs and a repetition rate of $\sim 80$ MHz. Additionally, an optical chopper system at $800$Hz is used to prevent sample heating while having a high pump intensity. The residual pump is removed with a spectral filter before collecting the PL emission in a spectrometer-CCD camera device. A complete description of the setup is presented in the Supplementary Note 2.

For the diffusion measurements, we used a continuous-wave (CW) laser. The rest of the optical measurement setup was similar.
%%%%%%%%%%%%%%%%%%%%%%%%%%%%%%%%%%%%%%%%%%%%%%%%%%%%%%%%%%%%%%%%%%%%%%%%%%%%%%%%%%%%%%%%%%%%%%%%%%%%%%%%%%%%%%%%%%%%%%%%%%%%%%
  \subsubsection{\bf Data availability}
The PL and diffusion data generated in this study have been deposited in the Figshare database under accession links:\\
 https://doi.org/10.6084/m9.figshare.25246012.v1
\\https://doi.org/10.6084/m9.figshare.25246006.v1
\\https://doi.org/10.6084/m9.figshare.25246009.v1
\\https://doi.org/10.6084/m9.figshare.25246015.v1
. 

\subsubsection{\bf Acknowledgements}
  We acknowledge fruitful discussions with N. Schine and A. Kollar. This work was supported by AFOSR FA95502010223, MURI FA9550-19-1-0399, FA9550-22-1-0339, NSF IMOD DMR-2019444, ARL W911NF1920181, and Simons and Minta Martin foundations.  Ming Xie is supported by Laboratory for Physical Sciences. R. Ni and Y. Zhou are supported by the U.S. Department of Energy, Office of Science, Office of Basic Energy Sciences Early Career Research Program under Award No. DE-SC-0022885.

\subsubsection{\bf Competing interests}
The authors declare no competing interests.

\subsubsection{\bf Contributions}
B. G., D.G.S.F, S.S., and M.H. conceived and designed the experiments. K.W. and T.T. supplied necessary material for the fabrication of the sample. B.G., D.S., and R.N. designed and fabricated the sample. J.V. and S.M. collaborated with the preparation of the setup at its initial stage. B.G., D.G.S.F., and S.S. performed the experiments. B.G., D.G.S.F., S.S., T.S.H., M.J.M., M.X., A.I., Y.Z., and M.H. analyzed the data and interpreted the results. T.S.H. and M.H. elaborated on the theoretical models presented in the manuscript. B.G., D.G.S.F., S.S., M.J.M., and M.H. wrote the manuscript, with input from all authors.

%%%%%%%%%%%%%%%%%%%%%%%%%%%%%%%%%%%%%%%%%%%%%%%%%%%%%%%%%%%%%%%%%%%%%%%%%%%%%%%%%%%%%%%%%%%%%%%%%%%%%%%%%%%%%%%%%%%%%%%%%%%%%%

\newpage
%\onecolumngrid

%%%%%%%%%% Merge with supplemental materials %%%%%%%%%%
%%%%%%%%%% Prefix a "S" to all equations, figures, tables and reset the counter %%%%%%%%%%
\setcounter{equation}{0}
\setcounter{figure}{0}
\setcounter{table}{0}
\setcounter{page}{1}
\makeatletter
\renewcommand{\theequation}{S\arabic{equation}}
\renewcommand{\thefigure}{S\arabic{figure}}
\pagenumbering{roman}
%%%%%%%%%% Prefix a "S" to all equations, figures, tables and reset the counter %%%%%%%%%%
%%%%%%%%%% Merge with supplemental materials %%%%%%%%%%

\noindent
\centering \textbf{\Large Supporting Information}\\

\twocolumngrid
\begin{flushleft}
\section{Contents}
{\bf 1. Device fabrication}
\\{\bf 2. Optical measurements}
\\{\bf 3. Electronic filling factor calibration}
\\{\bf 4. Determination of stacking angle via Second Harmonic Generation}
\\{\bf 5. Total emitted PL vs. Pump Intensity}
\\{\bf 6. Pulsed vs. CW excitation}
\\{\bf 7. Data analysis}
\\{\bf 8. Modeling X$_1$ and X$_2$}
\\{\bf 9. Diffusion measurements}
\\{\bf 10. System’s spectrum in the purely bosonic limit}
\\{\bf 11. Reproduction of the observations in Device D2}
\\{\bf 12. Comparison of U$_{ex-ex}$ and U$_{ex-e}$ with previously reported values}
\\{\bf 13. Optical signatures of photo-doping effect}
\end{flushleft}

%%%%%%%%%%%%%%%%%%%%%%%%%%%%%%%%%%%%%%%%%%%%%%%%%%%%%%%%%%%%%%%%%%%%%%%%%%%%%%%%%%%%%%%%%%%%%%%%%%%%%%%%%%%%%%%%%%%%%%%%%%%%%%
\justifying
\subsection{Supplementary Note 1. Device fabrication}

The $\rm{WSe}_2/\rm{WS}_2$ heterostructure shown in Supplementary Figure \ref{figS1} was fabricated using a dry-transfer method with a stamp made of a poly(bisphenol A carbonate) (PC) layer on polydimethylsiloxane (PDMS) \cite{DryTransfer}. All flakes were exfoliated from bulk crystals onto Si/SiO$_2$ (285 nm) and identified by their optical contrast. The top/bottom gates and TMD contact are made of few-layer graphene. The PC stamp and samples were heated to $60^{\circ}$C during the pick-up steps and released from the stamp to the substrate at $180^{\circ}$C. The PC residue on the device was removed in chloroform followed by a rinse in isopropyl alcohol and ozone clean. Sample transfer was performed in an argon-filled glovebox for improved interface quality. The electrodes consist of 3.5 nm of chromium and 70 nm of gold. They were fabricated using standard electron-beam lithography techniques and thermal evaporation.

\begin{figure}[htb!]
    \centering    
    \includegraphics[width=0.8\columnwidth]{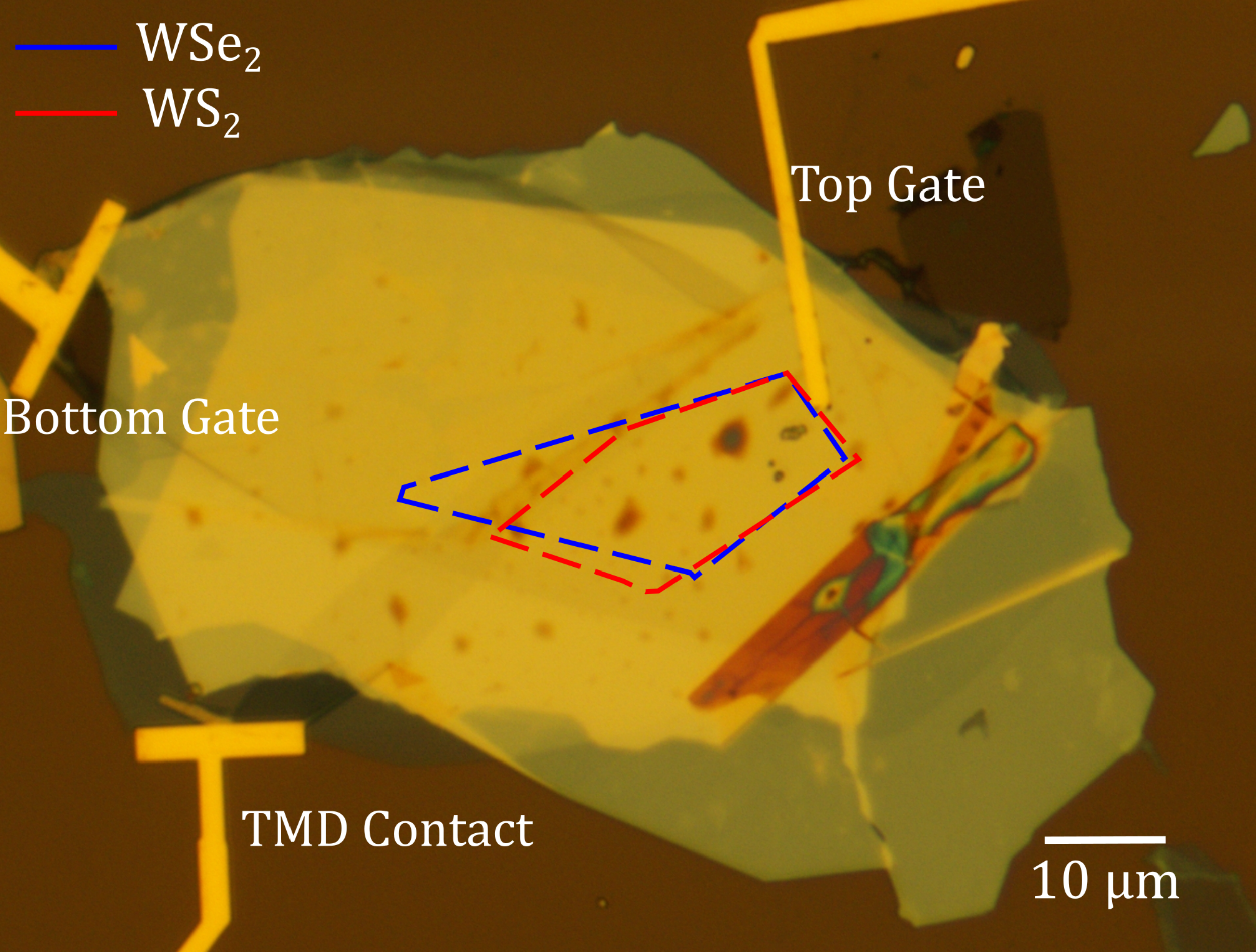}
    \caption{Microscope image of the $\text{WS}_2/\text{WSe}_2$ heterostructure device.}
    \label{figS1}
\end{figure}

%%%%%%%%%%%%%%%%%%%%%%%%%%%%%%%%%%%%%%%%%%%%%%%%%%%%%%%%%%%%%%%%%%%%%%%%%%%%%%%%%%%%%%%%%%%%%%%%%%%%%%%%%%%%%%%%%%%%%%%%%%%%%%
\begin{figure*}[htb!]
    \centering    
    \includegraphics[width=1.9\columnwidth]{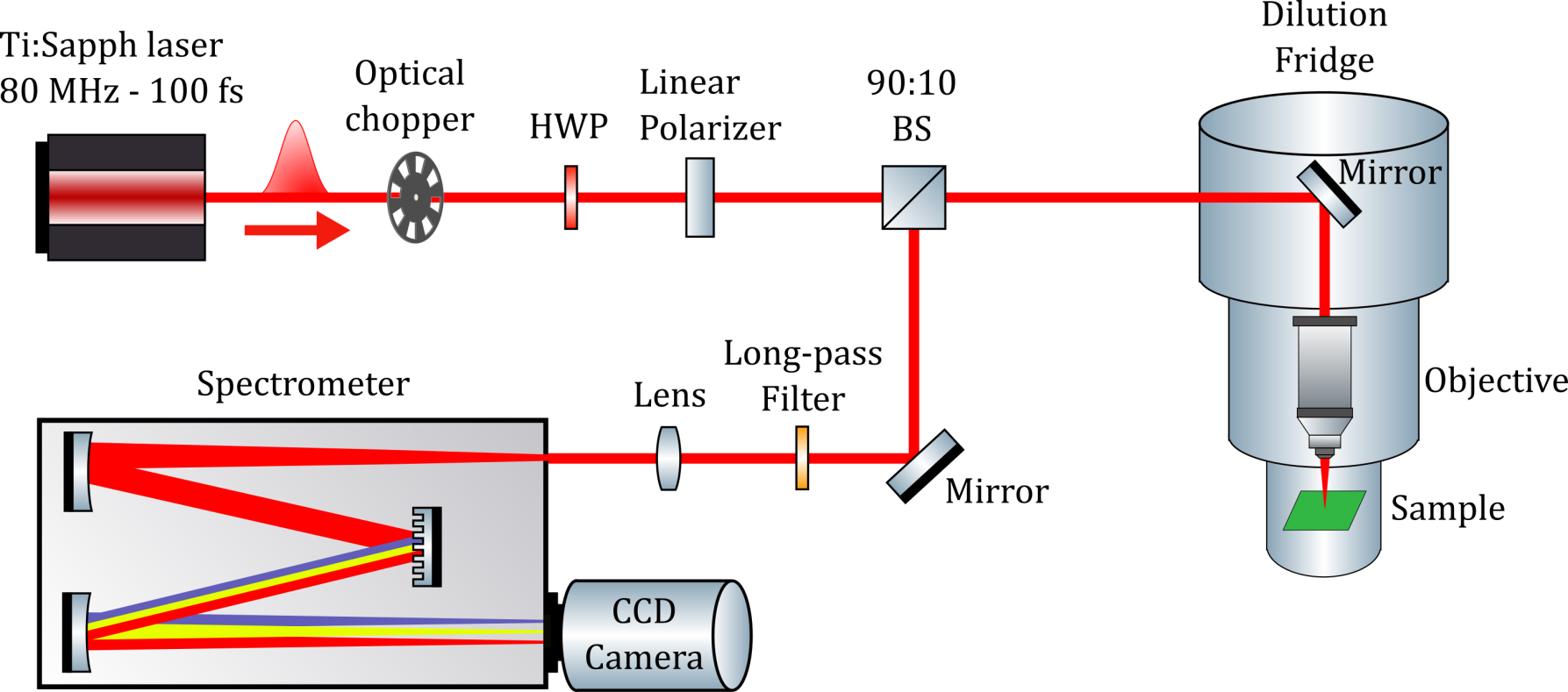}
    \caption{Schematic of the experimental setup. HWP$=$Half-wave plate, BS$=$Beam splitter}
    \label{fig_schematic}
\end{figure*}

\subsection{Supplementary Note 2. Optical measurements}

The sample is kept in a dilution refrigerator at a temperature of $3.5$K. For photoluminescence measurements, we use a confocal microscopy setup with an objective of magnification  $70\times$ and numerical aperture $\text{NA} = 0.82$. Our pumping source is a pulsed Ti:Sapphire laser tuned at $720$nm ($1.722$eV), with a pulse duration of $100$fs and a repetition rate of $\sim 80$ MHz. A half-waveplate placed before a polarizing beam-splitter is rotated to control the pump power. Additionally, an optical chopper system at $800$Hz is used to prevent sample heating while having a high pump intensity. A $750$nm long-pass filter was used to block the residual pump laser before collecting the PL emission in a spectrometer equipped with a $300$ grooves per mm diffraction grating and a CCD camera. The schematic of the setup is shown in Supplementary Figure \ref{fig_schematic}.

For the diffusion measurements, we used a continuous-wave (CW) Ti:Sapphire laser tuned at $708$nm. The rest of the optical measurement setup was similar. By applying a spatial filter to the reconstructed image of the diffusion pattern, we obtain spectral and spatial data of the exciton emission. This allows to image the spatial diffusion of each spectral component and extract the diffusion length.

%%%%%%%%%%%%%%%%%%%%%%%%%%%%%%%%%%%%%%%%%%%%%%%%%%%%%%%%%%%%%%%%%%%%%%%%%%%%%%%%%%%%%%%%%%%%%%%%%%%%%%%%%%%%%%%%%%%%%%%%%%%%%%

\subsection{Supplementary Note 3. Electronic filling factor calibration}

The electronic filling factor ($\nu_e$) can be estimated by combining the information about the crystalline structure of the bilayer device and a parallel capacitor model, as follows:

\textbf{Determination of moiré density $\mathbf{n_0}$:} In a purely fermionic Mott insulating state, the heterostructure will host one electron per moir\'e unit cell. The charge density in this case is given by ${\rm n}_0$, and it can be directly determined by the moiré periodicity through the relationship ${\rm n}_0 = 1/({L_M}^2 \sin \pi/3)$. Here, $L_M=a/\sqrt{\delta^2 + \theta^2}$ is the size of the moir\'e superlattice, $\delta = (a-a')/a \approx 4 \%$ is the lattice mismatch between $ \text{WSe}_2$ $ ( a = 0.328 \text{nm})$ and $\text{WS}_2$ $ (a' = 0.315 \text{nm})$, and $\theta $ is the twist angle between the two layers. Assuming $0^{\circ} \le \theta \le 1^{\circ}$, we obtain $1.72 \times 10^{12} \text{cm}^{-2}$ $\le {\rm n}_0 \le 2.04 \times 10^{12} \text{cm}^{-2}$.

 \begin{figure*}[htb!]
    \centering    
    \includegraphics[width=1.8\columnwidth]{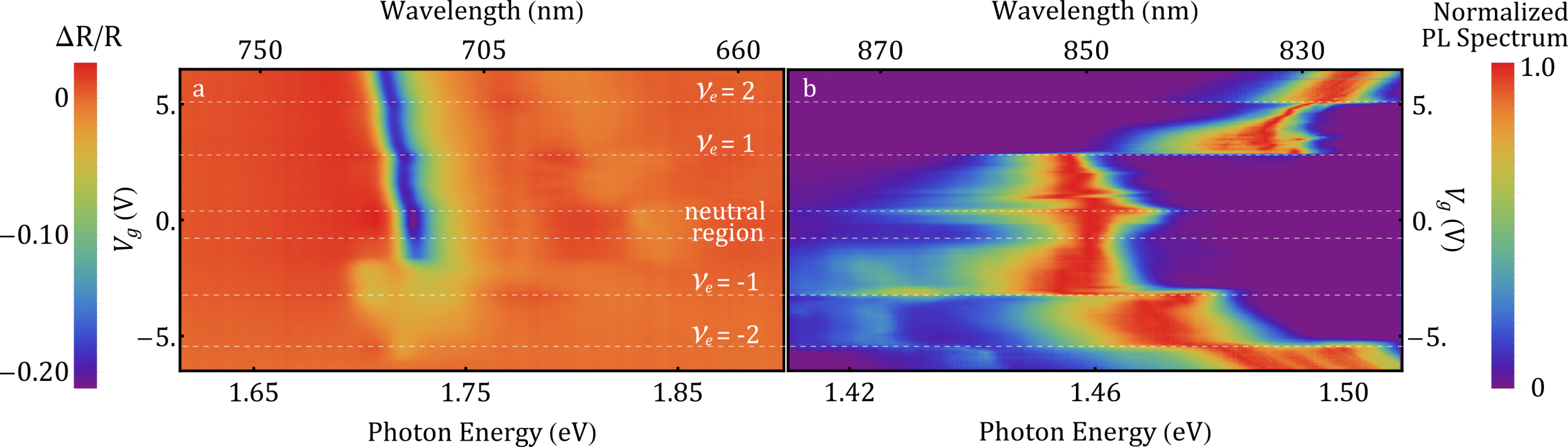}
    \caption{Reflection and photoluminescence spectra of our $\text{WS}_2/\text{WSe}_2$ heterobilayer device.}
    \label{figS2}
\end{figure*}

\textbf{Determination of electron density $\mathbf{n_{\rm e}}$:} From a parallel capacitor model, we can deduce the expression for the electron density ($n_{\rm e}$) in the bilayer. For a dual gate device, $n_{\rm e}$ is given by:
$$ n_{\rm e} = \frac{\epsilon_{\rm r} \epsilon_0\Delta V_{\rm g}}{d_{\rm t}}+\frac{\epsilon_{\rm r} \epsilon_0\Delta V_{\rm g}}{d_{\rm b}},$$ where $V_{\rm g}$ is the symmetrically applied gate voltage and $d_{\rm t} \approx d_{\rm b} \approx 40 $nm are the thicknesses of the top and bottom hBN dielectrics, respectively. They are determined from the optical contrast of the hBN flakes under the microscope. $\epsilon_0$ is the vacuum permittivity, and $\epsilon_r \approx 3$ is the relative permittivity of hBN \cite{Movva2017}. 
Having the moir\'e density $n_0$ and the electron density $n_{\rm e}$, the electronic (fermionic) filling factor can be expressed as $\nu_{\rm e}=n_{\rm e}/n_0$. From this model, we deduce that the gate voltage at which the electronic Mott insulator is established lies in the range $2.65$ V$<\!V_{\rm g}\!<\!3.04$~V. Where we are taking into account that the neutral region extends up to $V_{\rm g}\!=\!0.58$ V.
This estimation is in good agreement with our experimental data for reflectivity and PL. The results, displayed in Supplementary Figure \ref{figS2} show a consistent change in the optical response of the device at $V_\text{g}\!=\!2.98$ V. For the reflectivity (panel a), we observe a shift in the intralayer exciton energy. For the PL (panel b), a corresponding energy transition is observed for the interlayer exciton.

Although this calibration holds for both the hole-doped and the electron-doped sides, Supplementary Figure \ref{figS2}b shows an asymmetric behavior of the exciton interaction with the sign of the doping: the energy gap at $\nu_e=1$ on the electron doping side is larger than the one in the hole doping side. This is in agreement with theoretical predictions \cite{Naik2022} and experimental observations \cite{Xiong2023,Lian2023} that show a reduced value of the interaction strength due to the fact that holes and excitons reside in different high symmetry points of the moiré lattice. This observation indicates that electron doping is more suitable for the study of Fermi-Bose hybrid correlated states.

%%%%%%%%%%%%%%%%%%%%%%%%%%%%%%%%%%%%%%%%%%%%%%%%%%%%%%%%%%%%%%%%%%%%%%%%%%%%%%%%%%%%%%%%%%%%%%%%%%%%%%%%%%%%%%%%%%%%%%%%%%%%%%

\begin{figure*}
    \centering
    \includegraphics[width=1.5\columnwidth]{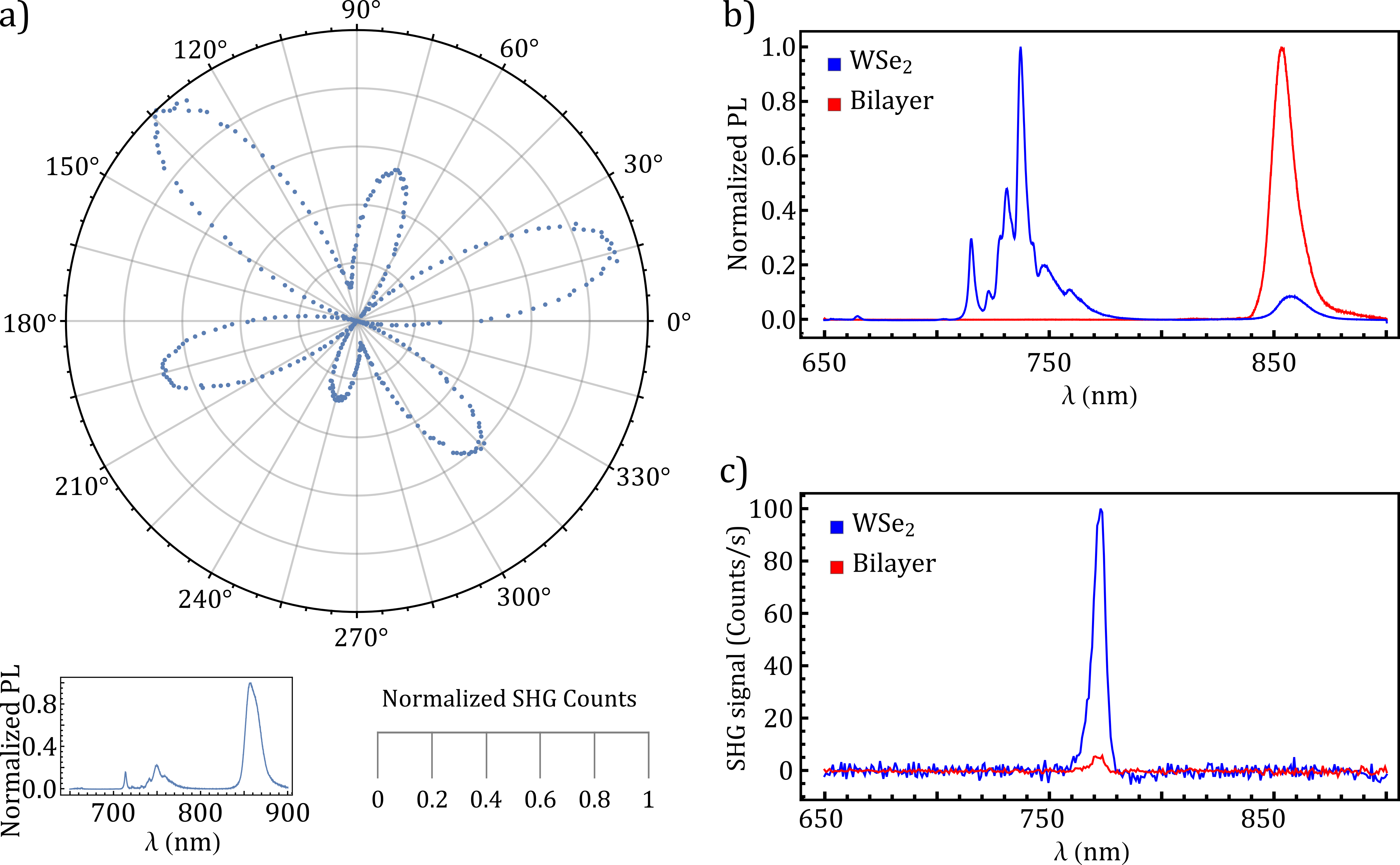}
    \caption{SHG sample characterization for the determination of the stacking angle. a) SHG angular dependence from a representative spot in the sample, characterized by the PL spectrum (inset). b) Characteristic PL emission spectra of spots on the sample with WSe$_2$ monolayer (blue) or WSe$_2$/WS$_2$ bilayer (red). c) SHG corresponding to the two spots of panel b. The strong suppression of the SHG emission efficiency indicates that the sample is stacked at an angle of 60$^{\circ}$.}
    \label{shg}
\end{figure*}

\subsection{Supplementary Note 4. Determination of stacking angle via Second Harmonic Generation}
We performed Second Harmonic Generation (SHG) measurements on our device in order to determine the stacking angle of the component monolayers. This is important because the system's properties vary for R-stacking and H-stacking. Using a 100 fs excitation at 850 nm with variable linear polarization, we are able to reconstruct the expected behavior of the SHG angular dependence for a crystalline structure with inversion symmetry breaking (Supplementary Figure \ref{shg}a).
We first identify spots on the sample with bilayer and monolayer regions by using the PL emission for the characterization (panel b). After identifying each spot, we proceed to measure the SHG spectrum while keeping the fs laser power constant at $\sim\!150 \mu W$. As demonstrated in ref.~\cite{Wang2019} and the Supplementary Material of ref.~\cite{Jin2021}, the stacking angle can generate a constructive (destructive) interference in the SHG spectrum for 0° (60$^{\circ}$) stacking. The strong suppression in the SHG efficiency observed in panel c, allows us to conclude that our sample corresponds to a bilayer with a 60$^{\circ}$ stacking angle (AB or H stacking).

%%%%%%%%%%%%%%%%%%%%%%%%%%%%%%%%%%%%%%%%%%%%%%%%%%%%%%%%%%%%%%%%%%%%%%%%%%%%%%%%%%%%%%%%%%%%%%%%%%%%%%%%%%%%%%%%%%%%%%%%%%%%%%

\subsection{Supplementary Note 5. Total emitted PL vs. Pump Intensity}
\begin{figure}[htb!]
    \centering    
    \includegraphics[width=\columnwidth]{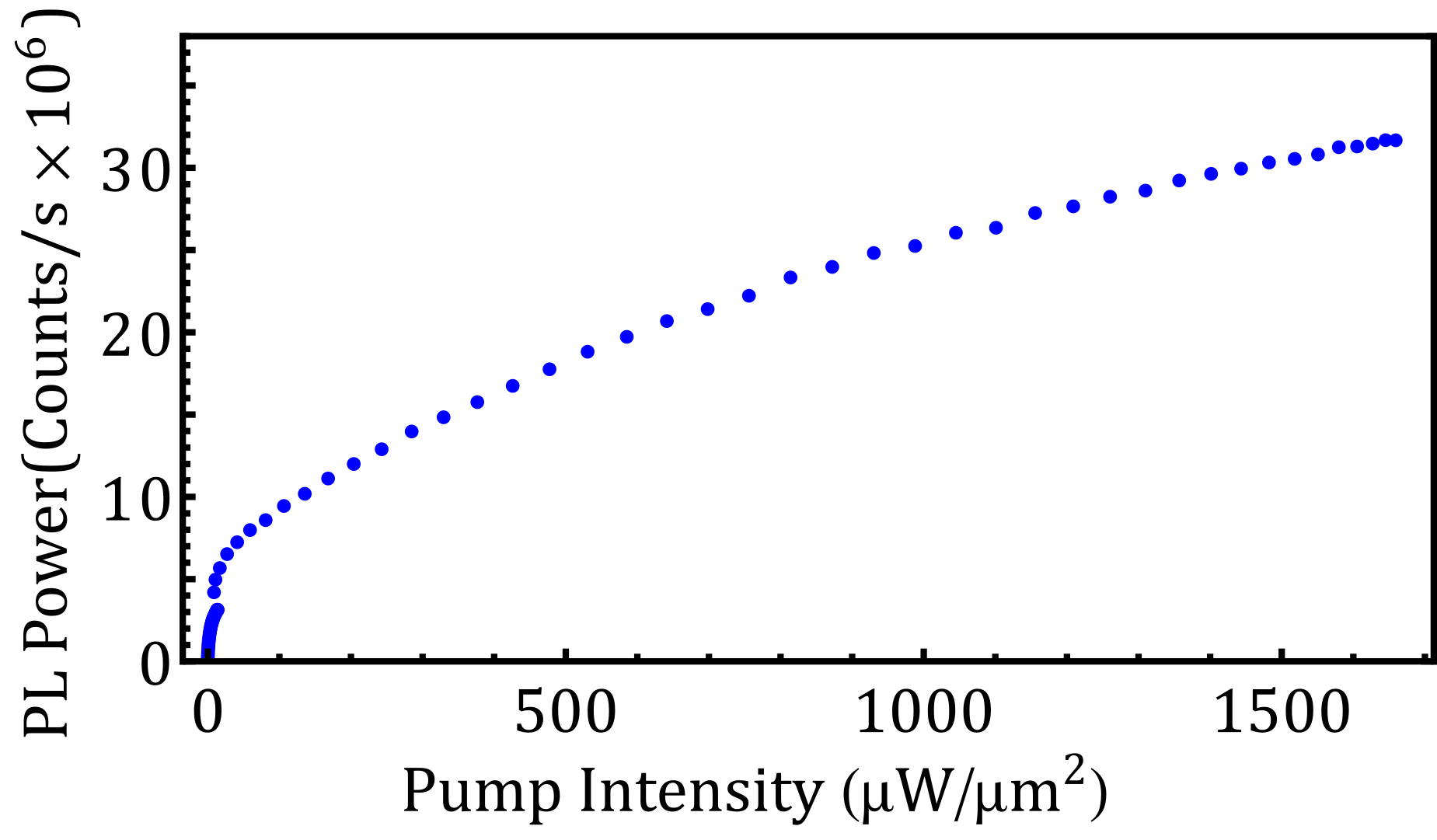}
    \caption{Pump intensity dependent total PL power at $V_\text{g}=0\text{V}$}
    \label{figS3}
\end{figure}

Due to the saturability of real semiconductor materials upon an intense optical pump, the laser intensity $I$ is not a suitable parameter to estimate the total exciton density in the bilayer. One can corroborate this from Supplementary Figure \ref{figS3}, by noticing that the emitted PL power does not follow a linear trend with increasing $I$. A more suitable quantity to monitor the changes in the excitonic density is the total emitted PL power. This quantity can be considered proportional to the total number of excitons formed in the structure, with the proviso that the radiative decay of the excitons does not change considerably within the phase space determined by $V_{\rm g}$ and $I$. For this reason, Figures 2 and 3 of the main text display the total PL power instead of the pump intensity. For simplicity, all the PL powers presented in the main text are normalized to a factor $10^6$.

%%%%%%%%%%%%%%%%%%%%%%%%%charge-filling%%%%%%%%%%%%%%%%%%%%%%%%%%%%%%%%%%%%%%%%%%%%%%%%%%%%%%%%%%%%%%%%%%%%%%%%%%%%%%%%%%%%%%%

\begin{figure*}
    \centering
    \includegraphics[width=2\columnwidth]{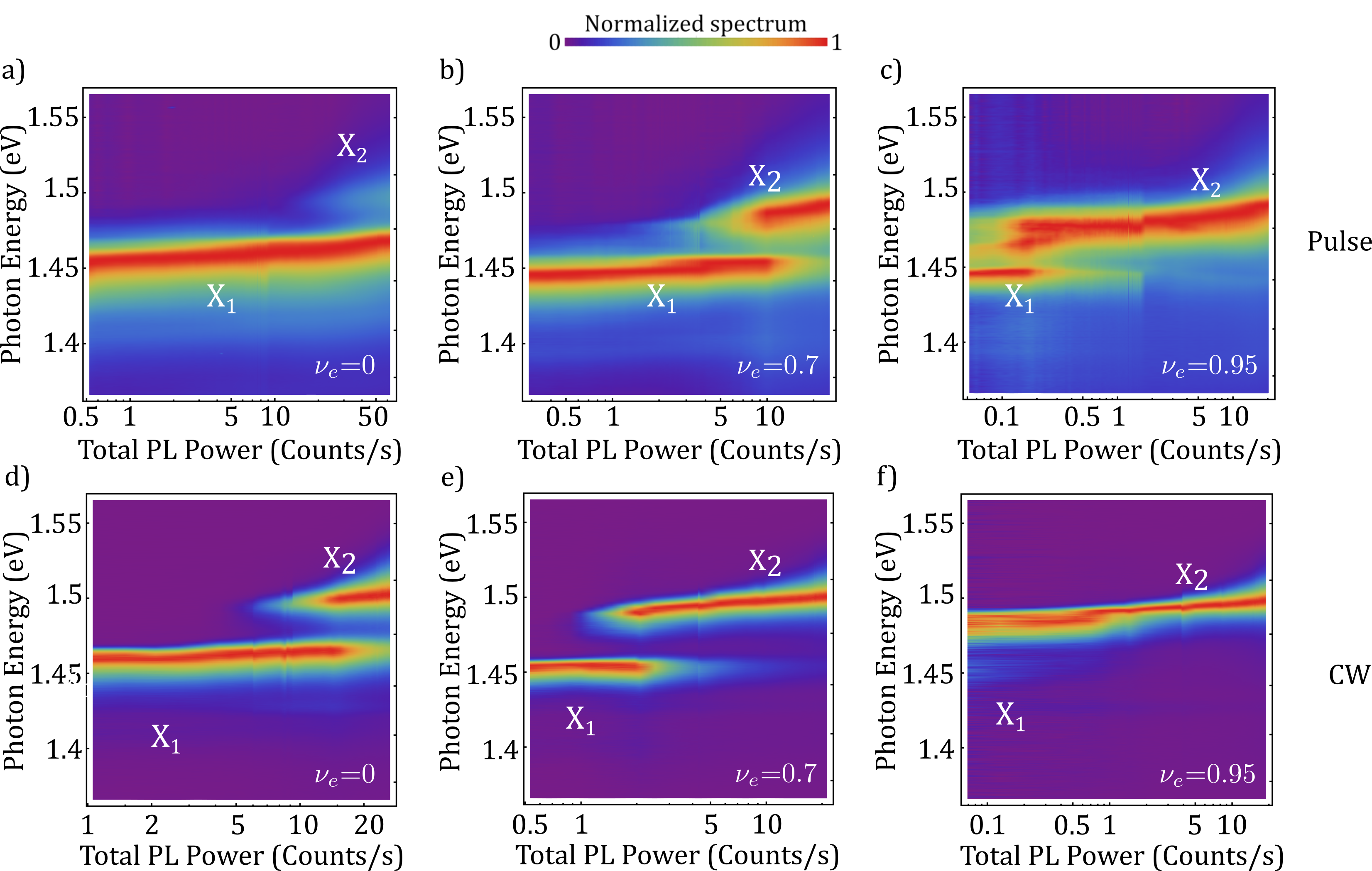}
    \caption{Normalized PL spectrum as a function of the total collected PL power for three different electronic filling factors under pulsed  (a-c) and CW (d-f) excitations. The peaks associated with single (X1) and double (X2) occupancies are indicated on each panel. The qualitative behavior remains unaltered under both excitation regimes.}
    \label{pulse-cw}
\end{figure*}

\subsection{Supplementary Note 6. Pulsed vs. CW excitation}
Although we performed experiments on both CW and pulsed regimes, the pulsed excitation has the important capability of reaching densely populated states while avoiding thermal effects, since the same populations can be achieved with much lower average power. Since Figs.~1, 2, and 5 of the main text present data collected in a pulsed excitation regime, it is important to verify that the high instant powers of the pulsed excitation are not inducing more complex nonlinear effects. This verification is done by comparing the data acquired in both excitation regimes. We use a Ti:Sapphire CW laser and perform the experiment in the same conditions. As shown in Supplementary Figure \ref{pulse-cw}, the qualitative behavior of the system is the same as in Supplementary Figure 2 of the main text: at low gate voltage only the single occupancy states are detected, and for increasing exciton density a secondary peak indicates the creation of double occupancies in the moiré lattice. The gate voltage dependence is also consistent: as one populates the lattice with electrons, the secondary peak becomes visible at lower gate voltages.

%%%%%%%%%%%%%%%%%%%%%%%%%%%%%%%%%%%%%%%%%%%%%%%%%%%%%%%%%%%%%%%%%%%%%%%%%%%%%%%%%%%%%%%%%%%%%%%%%%%%%%%%%%%%%%

\subsection{Supplementary Note 7. Data analysis}
\begin{figure}[htb]
    \centering    
    \includegraphics[width=\columnwidth]{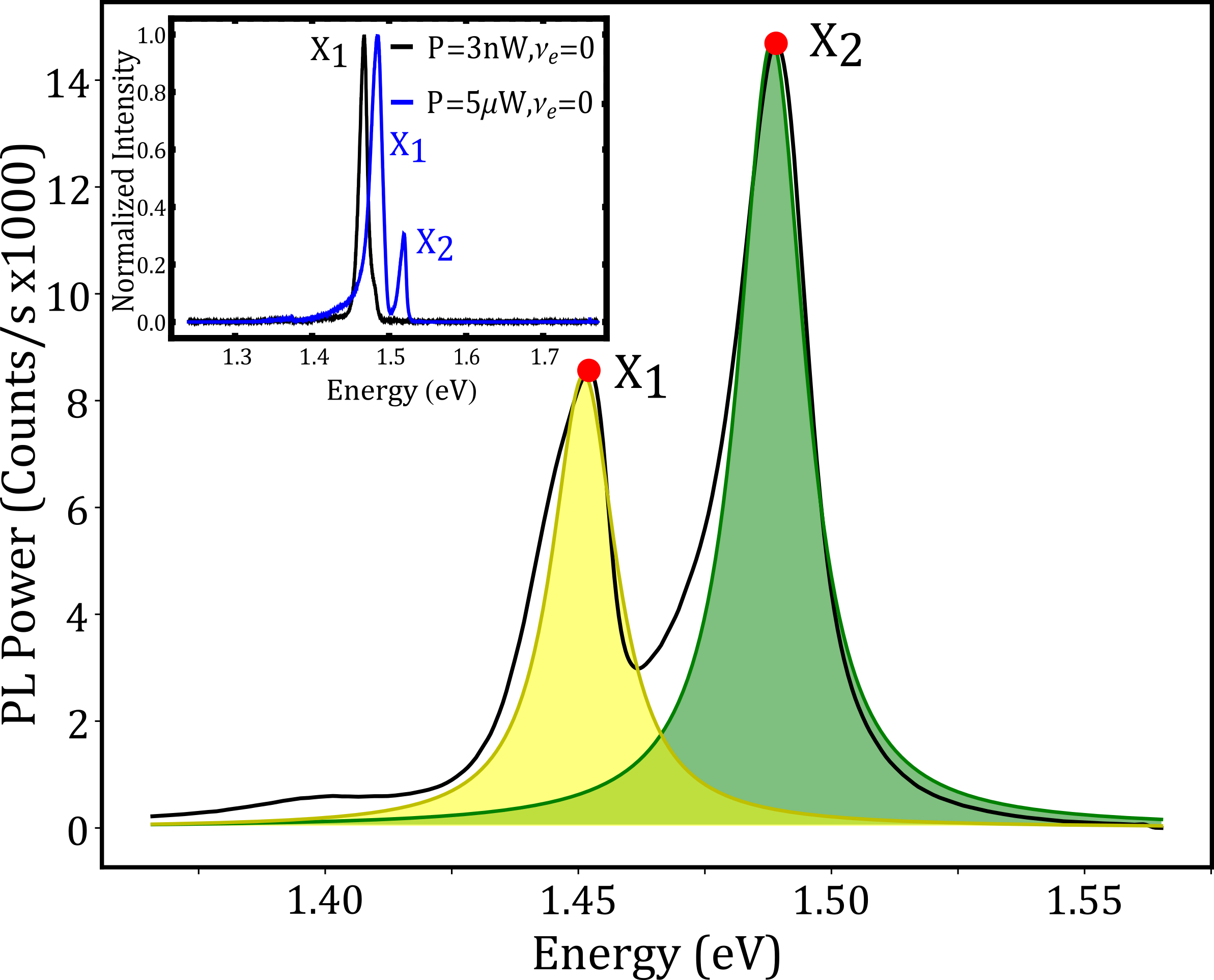}
    \caption{%\textcolor{red}{can we just delete this, it might give us trouble}
    A typical PL spectrum (V$_{\rm g}=2.1$ V and $I=2226$ $\mu$W/$\mu {\rm m}^2$) with the fitting functions obtained from the numerical method. Inset shows a wide PL spectrum at low and high pump powers with no WSe$_2$ intralayer excitons or trions.}
    \label{fit}
\end{figure}

To obtain relevant information about the collected spectra, we use a fitting routine. This procedure allows us to obtain the PL emission central energy, linewidth, and integrated PL power of each exciton line (X$_1$ and X$_2$). First, the PL spectra are processed by removing background noise and applying a low-pass Butterworth filter. After that, we extract the peaks from each spectrum by imposing constraints on the energy range, linewidth, prominence, and relative amplitude to the noise level. By setting a tolerance to the numerical error of the obtained values of intensity and energy of each peak, we fit each spectrum to a multi-Lorentzian distribution. Supplementary Figure \ref{fit} shows a representative spectrum with the corresponding fitting using two Lorentzian distributions.
From the fitting functions, we extract the central energies of X$_1$ and X$_2$ and integrate them over the obtained distributions to get the individual PL power. The Inset of  Supplementary Figure \ref{fit} shows a wide PL spectrum at low and high powers. At low power, we only observe X$_1$, while we see both X$_1$ and X$_2$ at high pump power. The process of creation of interlayer excitons is so efficient, that no PL is detected from intralayer excitons or trions. Hence, we only focus on the spectrum around the interlayer excitons for data analysis.

%%%%%%%%%%%%%%%%%%%%%%%%%%%%%%%%%%%%%%%%%%%%%%%%%%%%%%%%%%%%%%%%%%%%%%%%%%%%%%%%%%%%%%%%%%%%%%%%%%%%%%%%%%%%%%%%%%%%%%%%%%%%%%

\subsection{Supplementary Note 8. Modeling X$_1$ and X$_2$}
To obtain further insights into the physical mechanism dominating the system dynamics, we make a theoretical analysis using a 2-band model, as mentioned in the main text. Our goal is to deduce an analytical expression for the observed saturation behavior of the PL power from the X$_1$ population. This is achieved by considering the steady state of the system in a continuous-wave excitation regime. 
A population of electron-hole pairs $n_{\rm eh}$ is created by a Rabi-driving $\Omega(t)$. The charged plasma cascades down and relaxes to create $n_1$ 
singly occupied moir\'e sites (X$_1$) and $n_2$ doubly occupied sites (X$_2$). The formation of those states depends on the relaxation rate of the plasma ($\Gamma_{\text{relax}}$), the maximum number of available lattice sites ($n_{\rm max}(\nu_{\rm e}$)), and a $\beta$ factor, which determines the probability of the plasma to decay into the X$_1$ state. Under these conditions, the system's dynamics can be described by the set of equations:

\begin{subequations}
\begin{equation}
        \dv{n_{\text{eh}}}{t} = \eta \left| \Omega (t) \right|^2 -  \Gamma_{\text{relax}} n_{\text{eh}}
    %\dv{n_{\text{eh}}}{t} = -  \Gamma_{\text{relax}} n_{\text{eh}}^0,
\end{equation}
\begin{equation}
    \dv{n_1}{t} = \Gamma_{\text{relax}} \beta \dfrac{n_\text{max} (\nu_e) - n_1}{n_\text{max} (\nu_e)} n_\text{eh} - \Gamma_1 n_1
    \label{1b}
\end{equation}
\begin{equation}
    \dv{n_2}{t} = \Gamma_{\text{relax}} \left(1 - \beta \dfrac{n_\text{max} (\nu_e) - n_1}{n_\text{max} (\nu_e)} \right) n_\text{eh} - \Gamma_2 n_2
\end{equation}
\end{subequations}
\noindent where $\eta$ is the efficiency of the optical generation of electron-hole pairs and $\Gamma_1$ ($\Gamma_2$) is the decay rate of the excitonic states in a singly (doubly) occupied site. In a steady state condition, from Eq. \ref{1b}, we can obtain an analytical expression for $n_1$:

\begin{equation}
    n_1 = \dfrac{\Gamma_{\text{relax}} \beta n_\text{eh}}{\Gamma_1 + \Gamma_{\text{relax}} \beta \frac{n_\text{eh}}{n_\text{max}}}
    \label{eq_n1}
\end{equation}

In this regime $n_\text{eh}$ reaches an asymptotic value $n_\text{eh}\!=\!\frac{\eta}{\Gamma_{\text{relax}}} \left| \Omega \right|^2$. To associate this expression with our experimental data, we take into account that $n_{\rm eh}\Gamma_{\rm relax}$ is proportional to the total PL power and $n_1$ is proportional to the PL power ($P_1$) from the X$_1$ exciton band. After substituting variables we can conclude that the saturation behavior of X$_1$ should follow the functional form:
\begin{equation}
P_1=\text{P}_1^{\text{max}}\frac{{\rm P}}{\text{P}+\rm{P}_{\rm sat}}
\end{equation}
where $P_1^{\rm max}\!=\!n_{\rm max}$ is proportional to the PL power from X$_1$ in an ideal Mott insulating state and $P_{\rm sat}\!=\!\Gamma_1 n_{\rm max}/\beta$ determines the total PL emission at which X$_1$ saturates. This expression is used as a fitting model for the PL power from the X$_1$ states upon increasing total PL power, and provides a mathematical tool to estimate the bosonic occupation of the moir\'e lattice.
From this theoretical treatment, one can notice that the bosonic Mott insulating states are reached asymptotically. This means that $P_1$ can be arbitrarily close to $P_1^{\rm max}$ upon a high enough total PL power. In other words, the line that determines the Mott insulating phases in Figs.~1c, 2d, and 3d of the main text is mathematically not achievable. For this reason, the intensity denoted as $I^*$ in the referred figures, indicates the required pump intensity to drive the system into a state with a total population arbitrarily close to the lattice complete saturation.

%%%%%%%%%%%%%%%%%%%%%%%%%%%%%%%%%%%%%%%%%%%%%%%%%%%%%%%%%%%%%%%%%%%%%%%%%%%%%%%%%%%%%%%%%%%%%%%%%%%%%%%%%%%%%%%%%%%%%%%%%%%%%%

\begin{figure*}[htb!]
    \centering    
    \includegraphics[width=1.5\columnwidth]{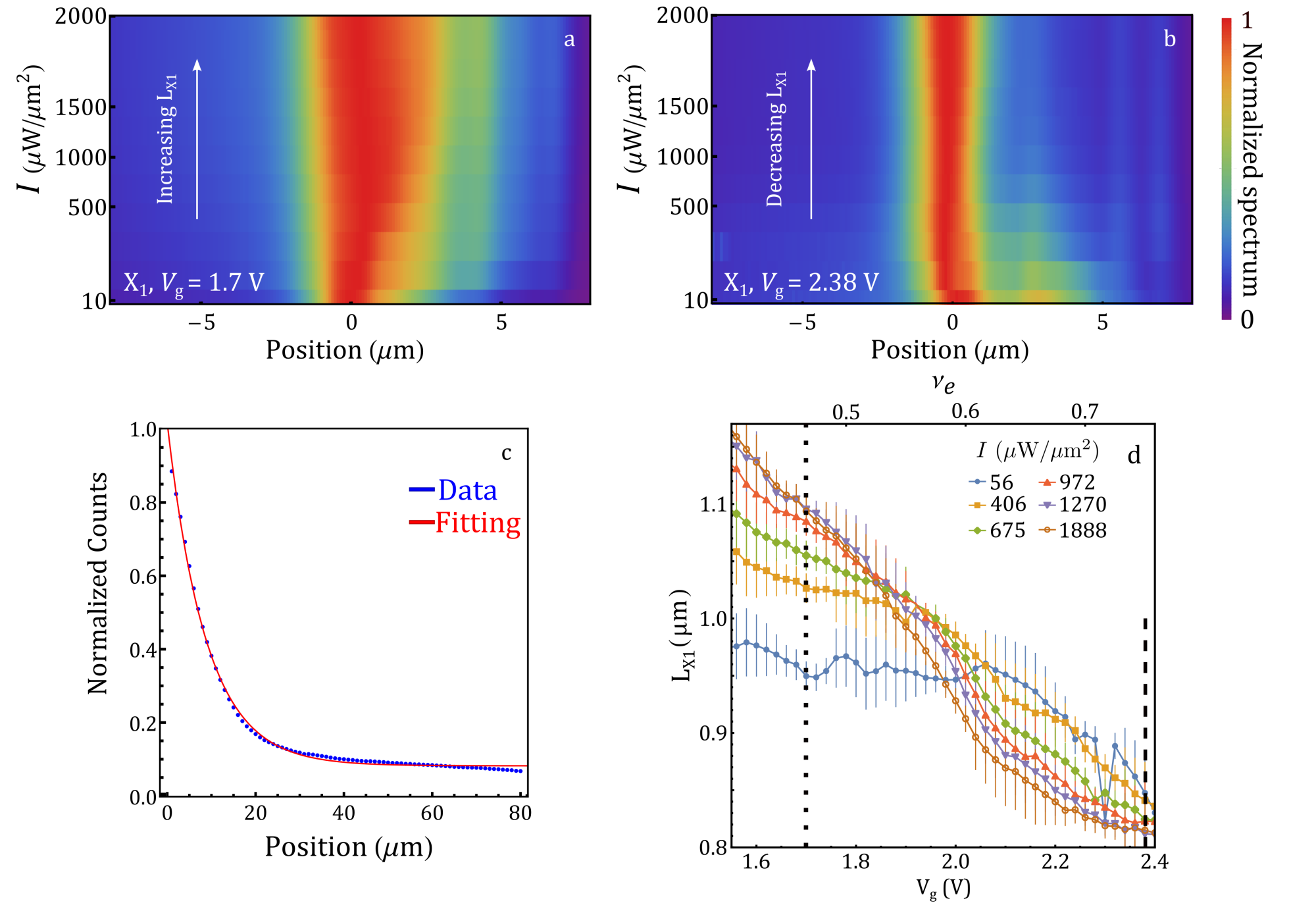}
    \caption{(a-b) Pump intensity dependence of the $\rm{X}_1$ spatial diffusion for $\nu_e \sim 0.47$ $ (\text{V}_g = 1.7 V)$ (a) and $\nu_e \sim 0.75$ $ (\text{V}_g = 2.38 V)$ (b). The panels show the inversion of the intensity dependence: from a diluted gas of bosons in panel a to a bosonic Mott insulator in panel b. (c) Exponential decay fitting of a typical $\rm{X}_1$ diffusion pattern. (d) $\rm{L}_{X1}$ for a reduced range of $V_{\rm g}$ to highlight the power dependence inversion. The error bars represent the standard errors for the diffusion length estimated from the exponential fitting.}
    \label{figS5}
\end{figure*}

\begin{figure*}[htb!]
    \centering    
    \includegraphics[width=1.9\columnwidth]{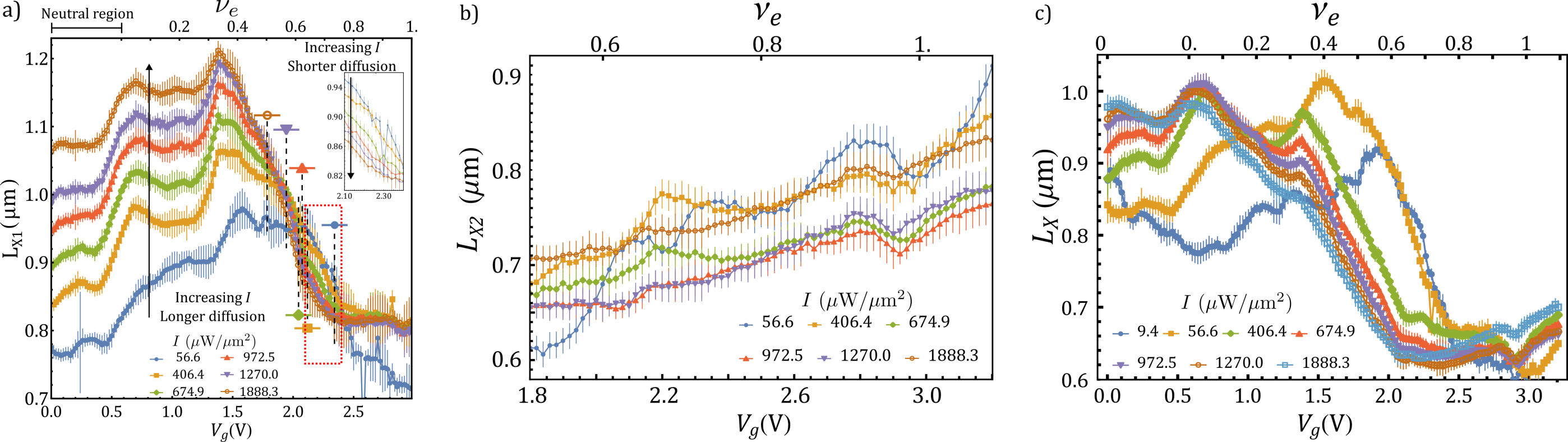}
    \caption{(a) $L_{X1}$ as a function of V$_{\rm g}$ for different pump intensities. In this panel, we include dashed lines corresponding to the crossing gate voltage at each power. They are a guide for the eye of where the incompressibility is expected to manifest at each power. (b)$\rm{L}_{X2}$ as a function of the gate voltage for a range of $\nu_e$ and for different I (c) $\rm{L}_{X}$ as a function of the gate voltage for a range of $\nu_e$ and different I. The error bars represent the standard errors for the diffusion length estimated from the exponential fitting.}
    \label{figS8}
\end{figure*}

\subsection{Supplementary Note 9. Diffusion measurements}

For the diffusion measurements, we use a continuous-wave laser tuned at 708 nm to create a steady population of excitons. By imaging the diffusion pattern with spectral resolution, we are capable of monitoring the diffusion properties of the X$_1$ and X$_2$ populations. We study the dependence of the diffusion length for each population as a function of $V_{\rm g}$ and $I$. To improve the spatial resolution, the image of the diffusion pattern is magnified $200$ times. We spectrally resolve the signals emitted from $\text{X}_1$ and $\text{X}_2$. Next, we fit each spatial diffusion profile to a function of the form $A\exp \left( - x/L_{\text{X}_i} \right)+b$, where $A$ is the PL power under the pumping laser spot, $x$ is the propagation distance, $L_{\text{X}_i}$ is the diffusion length of $\text{X}_i$ and $b$ is an offset that accounts for the base noise level. Supplementary Figure \ref{figS5} (a-b) shows the obtained power dependence for $L_{\text{X}_1}$ at $V_{\rm g}\!=\!1.7$ V and $V_{\rm g}\!=\!2.38$ V. We observe that $L_{\text{X}_1}$ increases with pump power at $1.7$ V but decreases at $2.38$ V. This inversion in the power dependence accounts for the formation of incompressible excitonic states. It is worth mentioning that the scale of the $I$ axis is not linear, because the power was modified by changing the polarization of the pump laser with a half-wave plate. Panel c shows the fitting subroutine used to extract the diffusion length of $\text{X}_1$. Panel d shows the measured value of $L_{\text{X}_1}$ for a range of $V_{\rm g}$ to highlight the inversion of the intensity dependence. The same analysis was performed for the diffusion of the X$_2$ states. The data (Supplementary Figure \ref{figS8}a) shows a shorter diffusion length, which makes it cumbersome to detect changes in $L_{\rm X2}$ due to the diffraction limit. However, in contrast to $L_{\rm X1}$, the $L_{\rm X2}$ increases with increasing $V_{\rm g}$, except for the reduction at $\nu_{\rm e}\sim1$, which can be attributed to the behavior of a population of excitons in presence of a fermionic Mott insulating state.   We finally analyze the diffusion data with no spectral resolution. For $\nu_e<1$, one can observe that the suppression of the diffusion takes place for lower $V_{\rm g}$ upon increasing pump intensity; a consequence of the bosonic saturation of the lattice due to the optical pump.

Delving deeper into the spatial diffusion, we identify a transfer of population from X$_2$ to X$_1$ as the excitons move away from the pump region. As observed in Supplementary Figure \ref{transfer}, for high pump power, the population of X$_1$ away from the excitation spot is larger than under it, a manifestation of the mentioned effect. At the excitation spot, the high exciton density leads to the formation of a localized Mott insulating state and therefore a large population of X$_2$. However, as the distance from the pumping region increases, part of the population transfers from X$_2$ to X$_1$. Notice how, for high excitation intensity, the X$_1$ signal becomes stronger away from the injection spot than under it. In this case, the X$_2$ population is acting as a reservoir for X$_1$ along the diffusion path. Importantly, this behavior does not change the interpretation of our results, but constitutes proof of the robustness of the effect; the population transfer from X$_2$ to X$_1$ would favor an apparently longer propagation constant for X$_1$, which, at most, would hinder the measured suppression of the diffusion.

\begin{figure}
    \centering
    \includegraphics[width=\columnwidth]{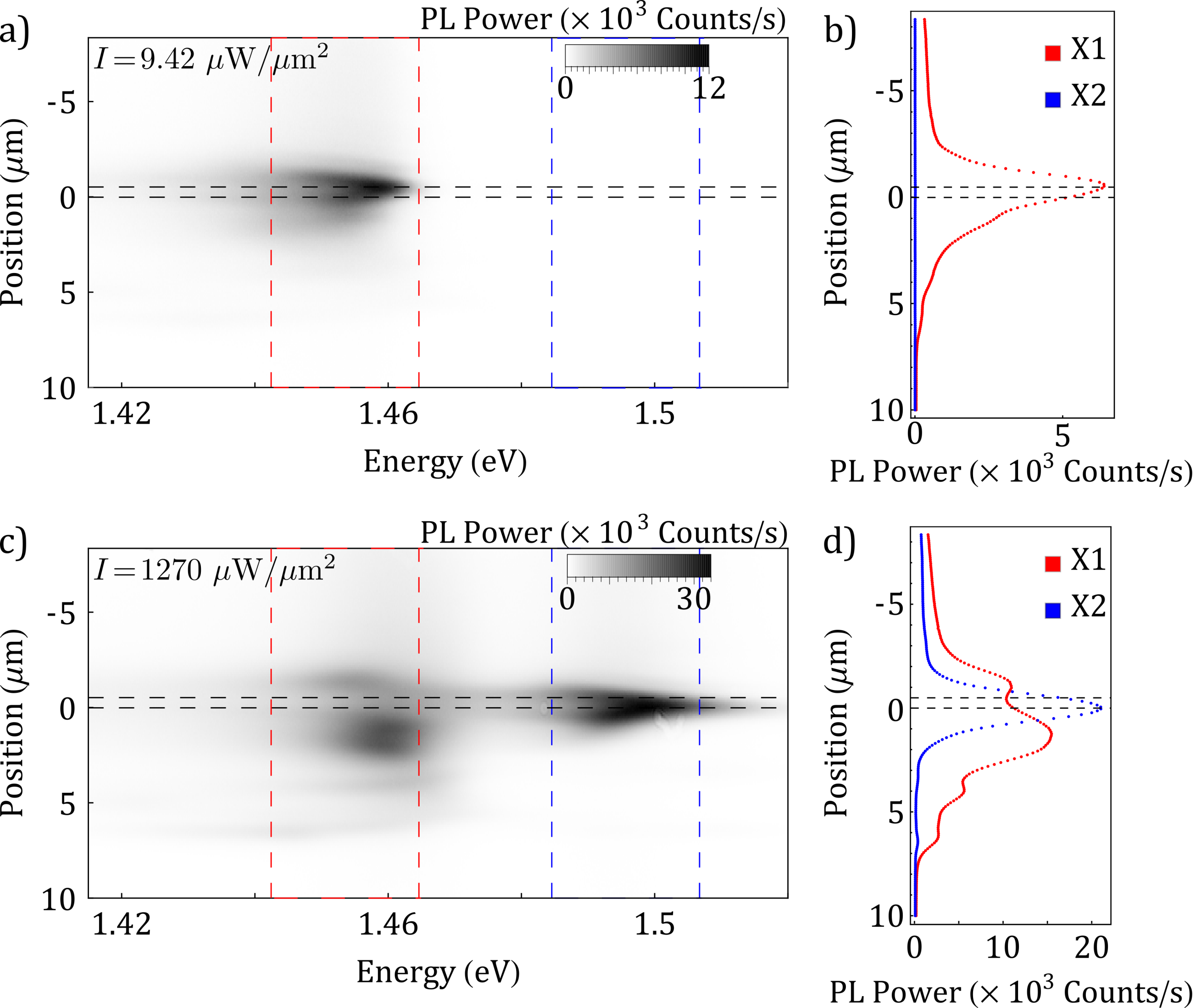}
    \caption{Diffusion of the system for $\nu_e\!=\!0$ and two different pumping intensities. a) 9.42 $\mu$W/$\mu$m$^2$ and c) 1270 $\mu$W/$\mu$m$^2$. The right panels (b and d) show the horizontally integrated intensity. The transfer of population from X$_2$ to X$_1$ becomes evident in panel d, where the X$_1$ intensity is larger outside the pumping spot due to a large reservoir of X$_2$ excitons that decay into X$_1$ as they propagate. }

    \label{transfer}
\end{figure}

\subsection{Supplementary Note 10. System's spectrum in the purely bosonic limit}
{In the absence of doping electrons, the PL spectrum displays a particular behavior for increasing excitonic population (Supplementary Figure 3a of main text): a blueshift at low occupancy and the emergence of a PL peak from a gapped state after a threshold PL power. In this ``reduced'' system, where the population is fully bosonic, we can model the system and account for the blueshift and the spectral gap by using the following Lindbladian master equation:
$$ \partial_t\hat{\rho}=-i[\hat{H},\hat{\rho}]+\sum_n\hat{\mathcal{L}}_n[\hat{\rho}], $$
where $\hat{H}$ corresponds to the Bose-Hubbard Hamiltonian:  
$$\hat{H}=\omega_{\rm x} \hat{x}^{\dagger}\hat{x}+\frac{U_{\rm ex-ex}}{2}\hat{x}^{\dagger}\hat{x}^{\dagger}\hat{x}\hat{x},$$
and the operators $\hat{x} (\hat{x}^{\dagger})$ stand for the annihilation (creation) of an exciton particle, $\omega_{\rm x}$ is their energy, and $U_{\rm ex-ex}$ is the on-site particle repulsion interaction energy. We consider two Lindbladian terms to take into account the laser pump and exciton loss. Specifically, the operator for channel $n$ is:
$$\hat{\mathcal{L}}_n[\hat{\rho}]=\hat{C}_n\hat{\rho}\hat{C}_n^\dagger-\frac{1}{2}\{\hat{C}_n^\dagger\hat{C}_n,\hat{\rho}\},$$
that includes the jump operators accounting for two incoherent processes: laser pump and exciton losses, with operators $\hat{C}_p$ and $\hat{C}_l$, respectively:

\begin{center}
$\hat{C}_{p}=\sqrt{\Gamma_{p}}\hat{x}^\dagger,$   \hspace{1.5cm}		$\hat{C}_{l}=\sqrt{\Gamma_{l}}\hat{x}.$\end{center}

Using the quantum regression theorem for the expected value $\langle\hat{x}^{\dagger}\hat{x}\rangle$, we calculate the spectral function and extract the central energy of the main peak. Given the set of theory data, we perform a routine for fitting the experimental results, obtaining the set:
\begin{center}
$U_{\rm ex-ex}\simeq32.4$ meV, \hspace{1.5cm}$\Gamma_l\simeq10.1$ meV, \hspace{1.5cm}	$\omega_{\rm x}\simeq1460$ meV
\end{center}}

\begin{figure}
    \centering
    \includegraphics[width=0.8\columnwidth]{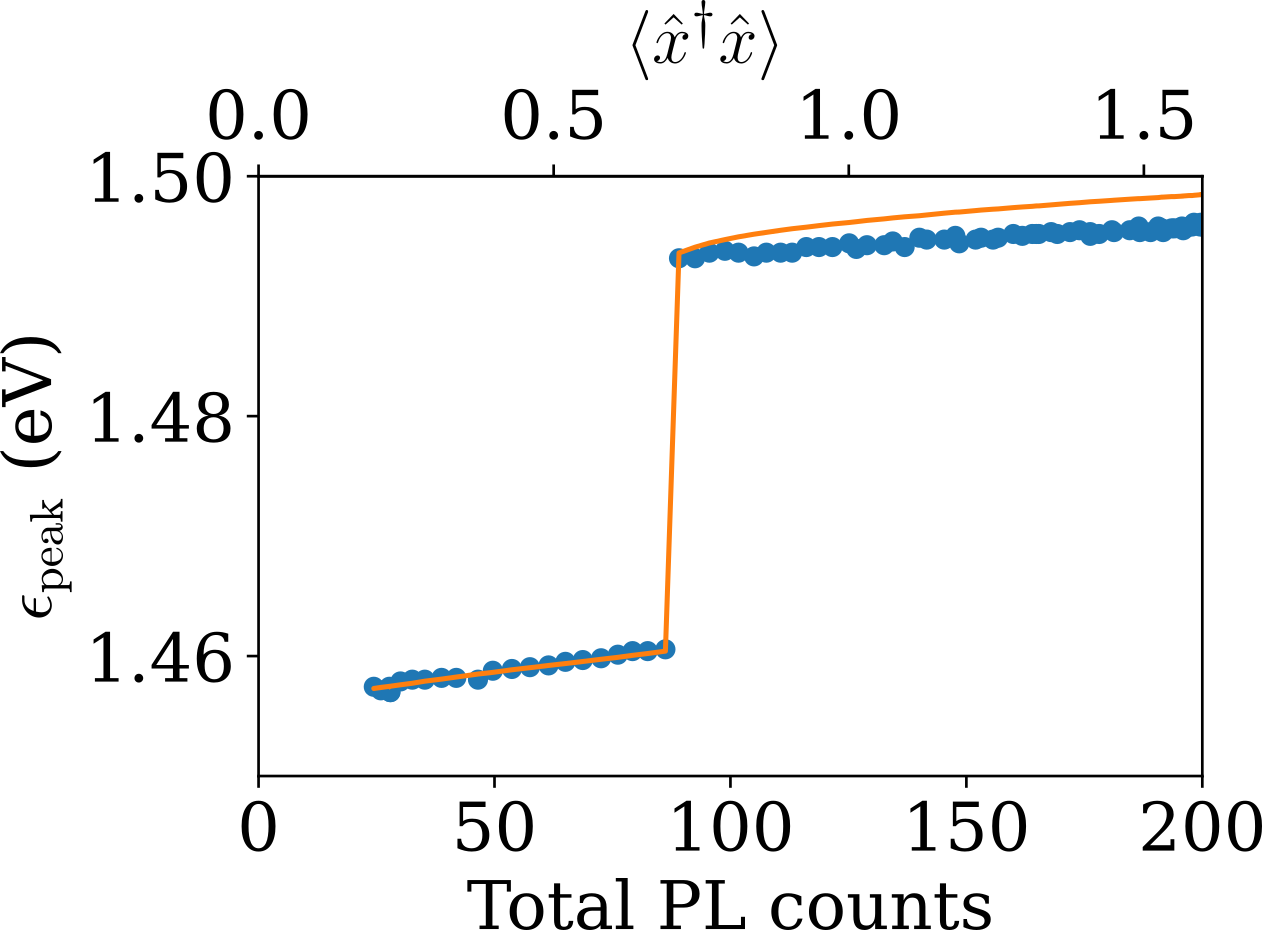}
    \caption{Continuous line: Main peak position of the calculated spectral function for a Bose-Hubbard model under the Master equation formalism. The blue circles correspond to the obtained experimental data (from Fig.~3a of the main text) for the PL power dependence of the spectrum at $\nu_e\!=\!0$.}    \label{hubbard}
\end{figure}

The results, displayed in Supplementary Figure \ref{hubbard} show a very good agreement with the experimental results. This figure provides important insights into the nature of U$_{\rm ex-ex}$. It shows that the on-site exciton repulsion leads to a spectral blueshift even before the system reaches an insulating state. Importantly, such a master equation cannot explain the intensity-dependent behavior of X$_1$. A unified model including higher-order jump operators (and hence more fitting parameters) may be required. We anticipate this master equation description and the rate equations presented in section VIII being simplifications of a unified theory, which needs further investigation.

%%%%%%%%%%%%%%%%%%%%%%%%%%%%%%%%%%%%%%%%%%%%%%%%%%%%%%%%%%%%%%%%%%%%%%%%%%%%%%%%%%%%%%%%%%%%%%%%%%%%%%%%%%%%%%%%%%%%%%%%%%%%%%

\begin{figure}
    \centering    
    \includegraphics[width=\columnwidth]{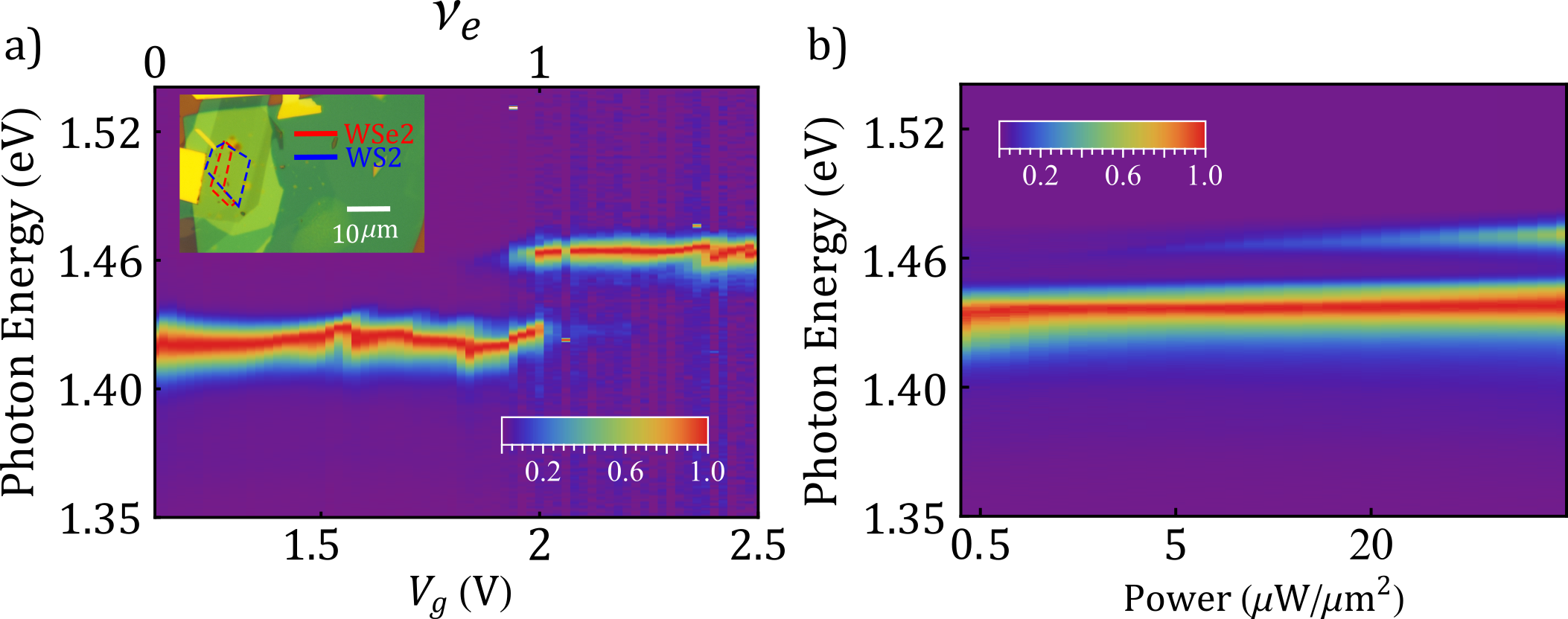}
    \caption{(a) Gate voltage-dependent normalized PL in device D2 for a pump intensity of 0.048 $\mu$W/$\mu m^2$. The splitting between X$_1$ and X$_2$ (U$_{\rm{ex-e}}$) is 34 meV. Inset shows an optical image of device D2.} (b) Power-dependent normalized PL in device D2 at $\nu_e=0$. The splitting between X$_1$ and X$_2$ (U$_{\rm{ex-ex}}$) is 35 meV.
    \label{device2}
\end{figure}

\begin{figure}[htb!]
    \centering    
    \includegraphics[width=\columnwidth]{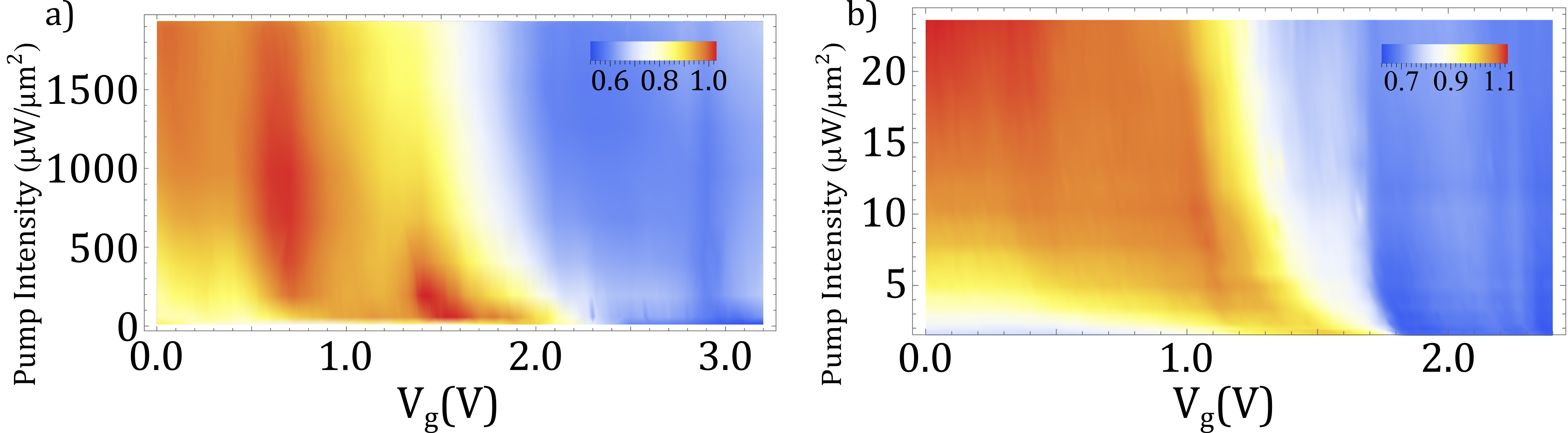}
    \caption{Power and gate voltage-dependent diffusion length of X$_{\rm{tot}}$ in device D1 (same as the device in the main text) and device D2. The difference in the magnitude of the power range is because the exciton lifetime of D2 is about 30 times longer than D1.}
    \label{diffusion3d}
\end{figure}

\subsection{Supplementary Note 11. Reproduction of the observations in Device D2}

Further verification of our experimental results is done by fabricating a second identical device (D2), whose picture can be observed in the inset of Supplementary Figure \ref{device2}a. We perform electron doping-dependent and power-dependent PL using a CW source. The results, displayed in Supplementary Figure \ref{device2}a-b are in agreement with the observations reported in the main text.

We also measured the diffusion length in the new device and compared the results with the ones from device D1. In Supplementary Figure \ref{diffusion3d}, we present the diffusion length in devices D1 (same as the one in the main text) and D2 when varying both the electronic filling factor and pump intensity. As it can be observed, in both devices there is a range of gate voltages for which increasing power brings a reduction of the diffusion length, confirming the formation of excitonic correlated states in both structures.

\begin{figure}[htb!]
    \centering    
    \includegraphics[width=\columnwidth]{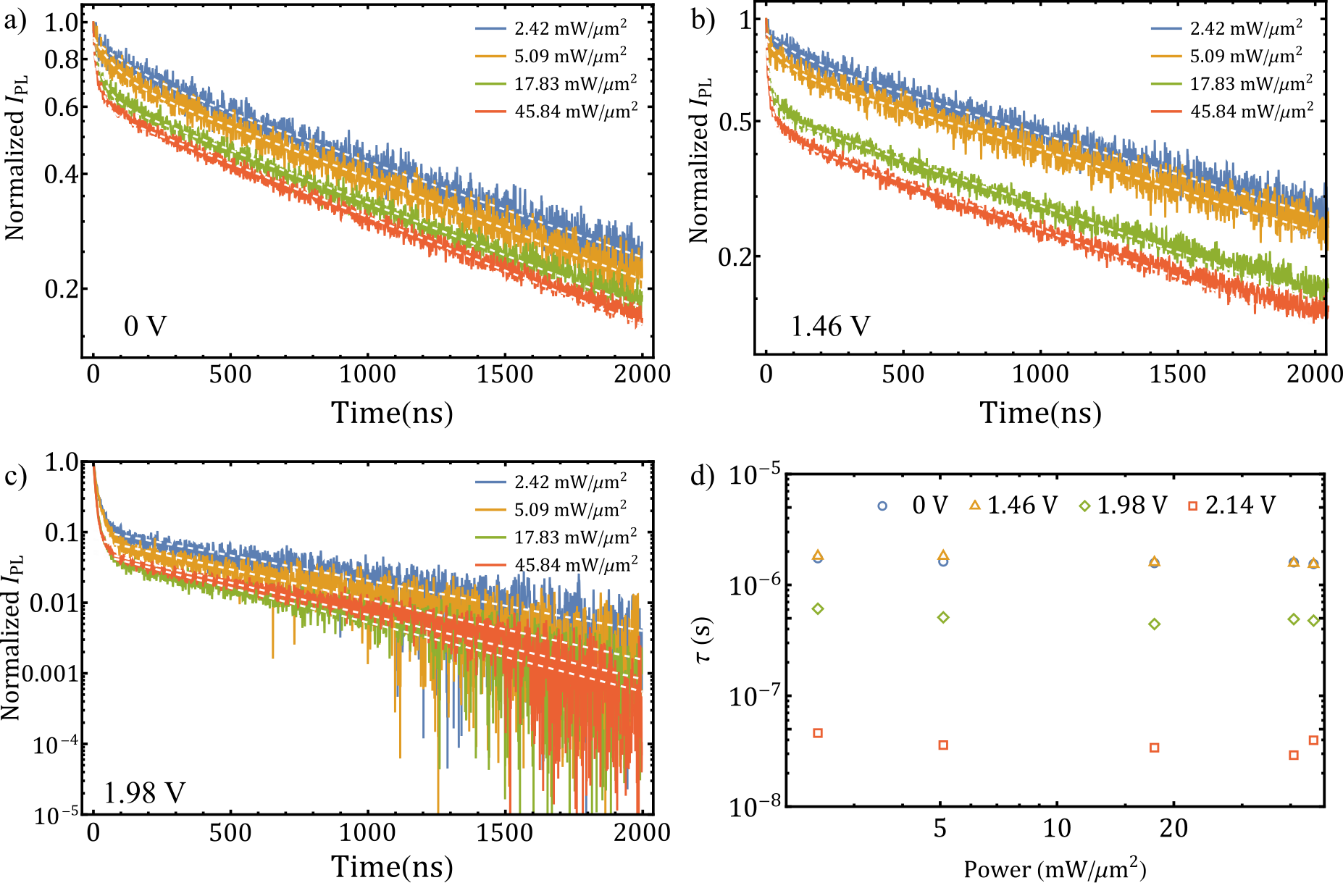}
    \caption{Time-dependent PL normalized at t = 0 for different peak powers at 0V (a), 1.46V (b), and 1.98V (c). For this measurement, we use a pulsed laser with a 500-kHz repetition rate and 100 ps “on” time. The power we quoted is the peak power. Dashed white lines correspond to double exponential fits. (d) Long lifetime values at different powers and different gate voltages.}
    \label{lifetime}
\end{figure}

Supplementary Figure \ref{lifetime} presents the time-resolved PL in D2 for different powers and doping levels. Panel d shows a noticeable decrease in the lifetime with increasing gate voltage (electron doping). However, with increasing power, we observe no significant change in lifetime but only the emergence of a fast-decaying population. The existence of two decay times has been reported in both multilayer and monolayer TMD systems \cite{Jiang2021}. In this device, we detect a fast-decaying population two orders of magnitude lower than the population with long decay times.

%%%%%%%%%%%%%%%%%%%%%%%%%%%%%%%%%%%%%%%%%%%%%%%%%%%%%%%%%%%%%%%%%%%%%%%%%%%%%%%%%%%%%%%%%%%%%%%%%%%%%%%%%%%%%%%%%%%%%%%%%%%%%%

\subsection{Supplementary Note 12. Comparison of U$_{\rm{ex-ex}}$ and U$_{\rm{ex-e}}$ with previously reported values}

\begin{table}
\caption{Compilation of reported values of the interaction energies U$_{\rm{ex-e}}$ and U$_{\rm{ex-ex}}$ in the literature.\label{tab}}
\begin{tabular}{ |p{0.25\columnwidth}||p{0.2\columnwidth}|p{0.2\columnwidth}|p{0.2\columnwidth}|}
 \hline
 & U$_{\rm{ex-e}}$ (meV) & U$_{\rm{ex-ex}}$ (meV) & $\dfrac{\rm{U}_{\rm{ex-e}}}{\rm{U}_{\rm{ex-ex}}}$ \\
 \hline \hline
 Device D1   & 27  &  32 &  0.84 \\ \hline
 Device D2 &  34 & 35 & 1  \\ \hline
 Park \textit{et al.} (R stack) \cite{Park2023} & - & 30-37 & - \\ \hline
 Lian \textit{et al.} (H stack) \cite{Lian2023} & 41 & 32 & 1.28 \\ \hline
 Lian \textit{et al.} (R stack) \cite{Lian2023} &  17  &  44 &  0.39 \\ \hline
 Xiong \textit{et al.} (H stack) \cite{Xiong2023} & 35 & 15 & 2.3  \\ 
 \hline
\end{tabular}
\end{table}

The exciton-exciton interaction energy U$_{\rm{ex-ex}}$ and exciton-electron interaction energy U$_{\rm{ex-e}}$ have been recently reported in the literature. Table \ref{tab} compiles them along with the values we measured in our devices. It is important to note the variations seen in the values of interaction energies even in similar heterostructure architectures. Although most of the interaction energies reported are in the range $\sim 15-44$ meV, no clear trend has been observed in experimentally reported values of $\rm{U}_{\rm{ex-e}}/\rm{U}_{\rm{ex-ex}}$.

\begin{figure}[htb!]
    \centering    
\includegraphics[width=\columnwidth]{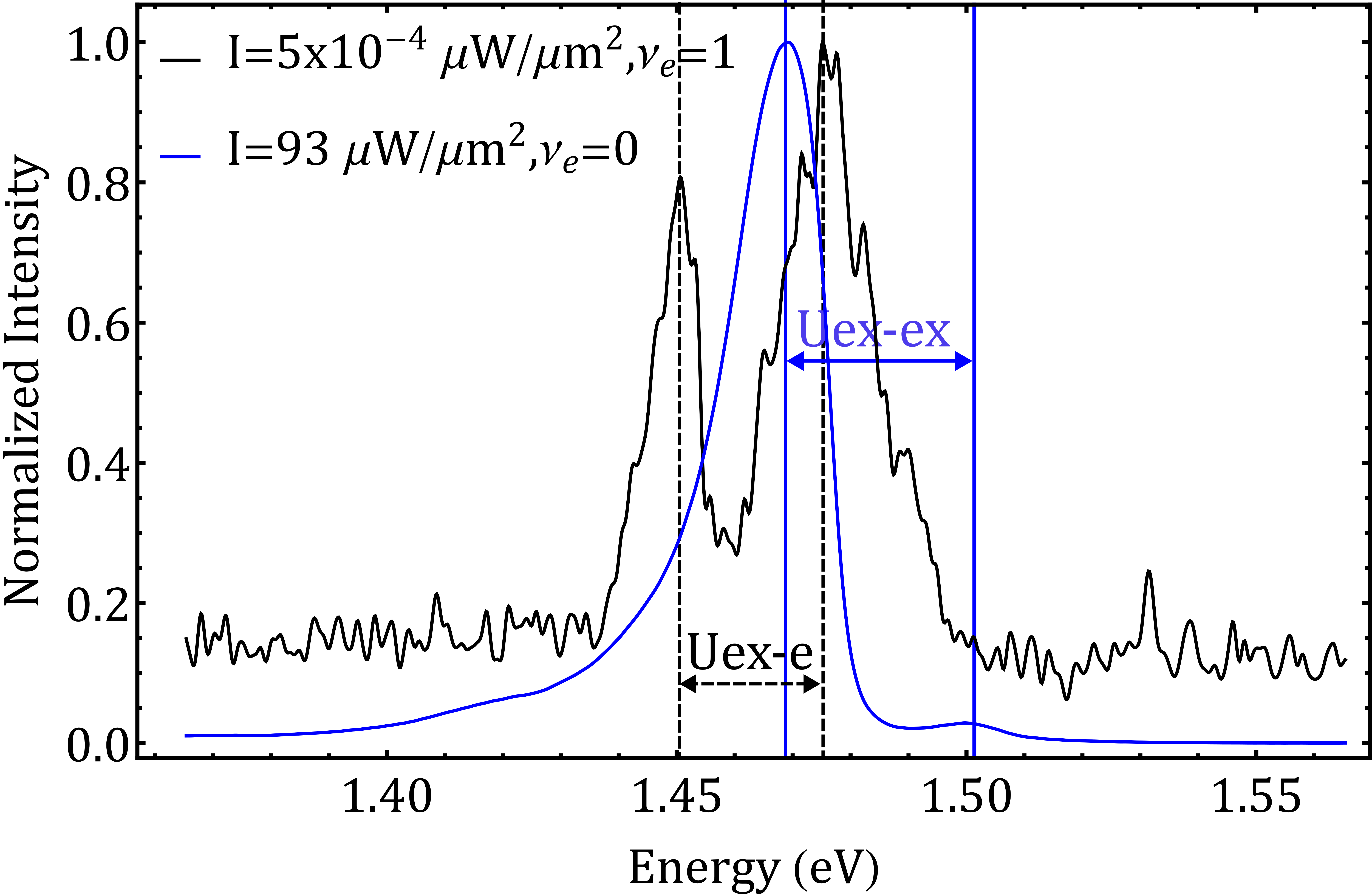}
    \caption{PL spectrum of device D1 showing $\rm{U}_{\rm{ex-e}} \sim 27$ meV and $\rm{U}_{\rm{ex-ex}} \sim 32$ meV.}
    \label{U}
\end{figure}

Supplementary Figure \ref{U} shows how we extracted values of interaction energies from the PL spectra. The energy gap between X$_1$ and X$_2$ at low power and $\nu_e=1$ gives us the exciton-electron interaction energy U$_{\rm{ex-e}}$, and the energy gap between X$_1$ and X$_2$ at lowest power for which X$_2$ appears even at $\nu_e=0$ gives us the exciton-exciton interaction energy U$_{\rm{ex-ex}}$.\\

%%%%%%%%%%%%%%%%%%%%%%%%%%%%%%%%%%%%%%%%%%%%%%%%%%%%%%%%%%%%%%%%%%%%%%%%%%%%%%%%%%%%%%%%%%%%%%%%%%%%%%%%%%%%%%%%%%%%%%%%%%%%%%

\begin{figure}[htb!]
    \centering    
    \includegraphics[width=\columnwidth]{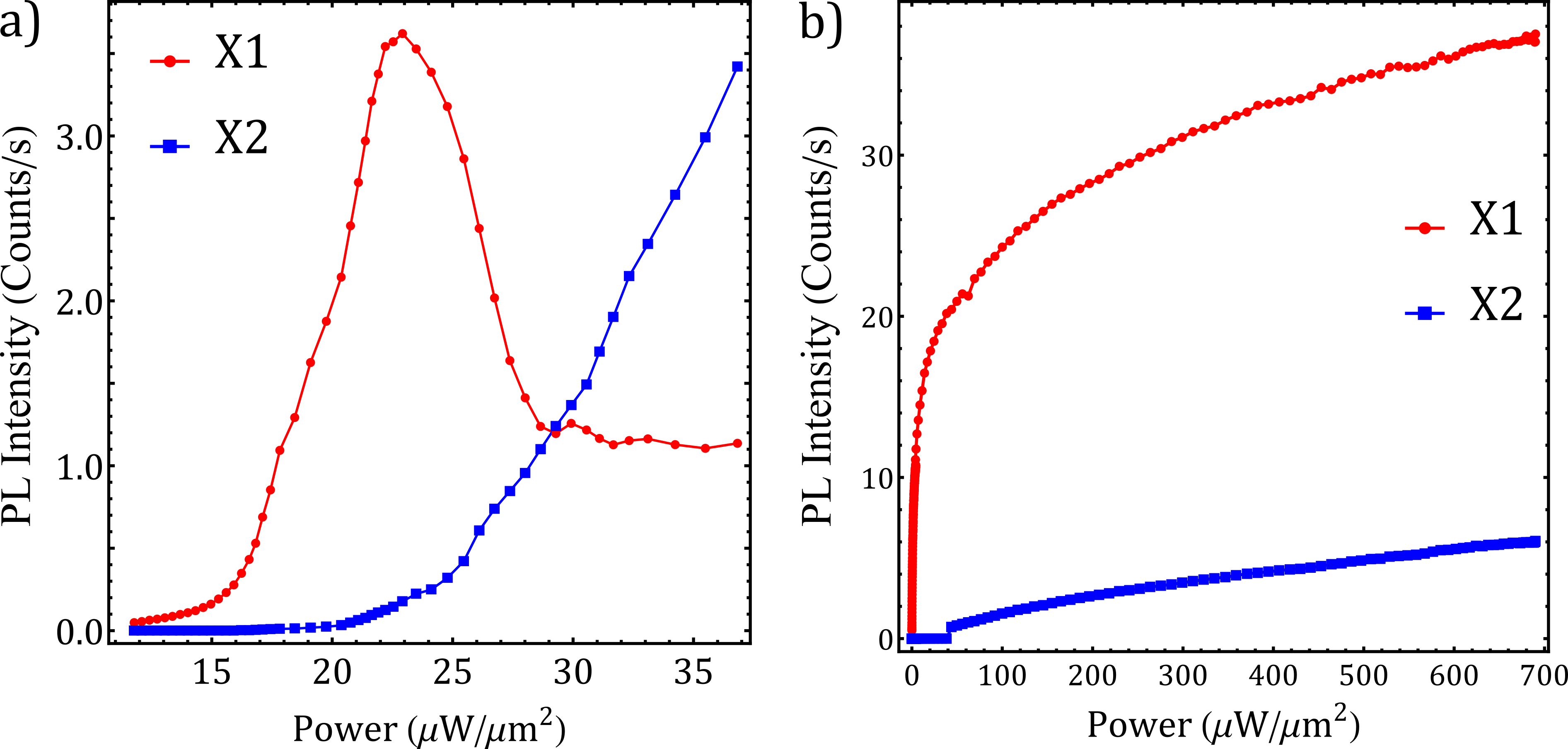}
    \caption{Evolution of PL intensity for X$_1$ (red) and X$_2$ (blue) as a function of the total excitation intensity for two different excitation regimes: a) device D2 pumped with an out-of-resonant CW laser (633nm) and b) device D1 pumped resonantly (720nm). Due to the photo-doping effect, the intensity of X$_1$ reduces after a threshold power, in contrast to the resonant excitation, where the PL intensity increases monotonically. In both panels $\nu_e=0.3$. Like in the main text, the presented counts are divided by a factor $10^6$.
}
    \label{photodoping}
\end{figure}

\subsection{Supplementary Note 13. Optical signatures of photo-doping effect}

Under the strong pumping power regime, one may expect the existence of a photodoping effect in the system. To alleviate this concern, we perform pump power-dependent measurements under off-resonant and on-resonant cases, which show the existence of a photodoping effect, and no photo-doping effect in the system, respectively. Supplementary Figure \ref{photodoping}a presents the X$_1$ and X$_2$ energy evolution of device D2 under a non-resonant pump (633 nm), where we can observe a strong photo-doping effect - a reduction of X$_1$ intensity as a function of pumping power, consistent to the gate voltage-dependent PL spectrum at low power (Fig.~2e in the main text). However, while using the resonant pump (720 nm) for D1, as indicated in panel b, the X$_1$ emission monotonically increases for the whole range of pump power, an observation confirms that no photo-doping effect exists in the system.
%%%%%%%%%%%%%%%%%%%%%%%%%%%%%%%%%%%%%%%%%%%%%%%%%%%%%%%%%%%%%%%%%%%%%%%%%%%%%%%%%%%%%%%%%%%%%%%%%%%%%%%%%%%%%%%%%%%%%%%%%%%%%%
\bibliography{biblio}

\end{document}